\pdfoutput=1

\documentclass[twoside,twocolumn,9pt]{article}
\usepackage{extsizes}
\usepackage[super,sort&compress,comma]{natbib} 
\usepackage[version=3]{mhchem}
\usepackage[left=1.5cm, right=1.5cm, top=1.785cm, bottom=2.0cm]{geometry}
\usepackage{balance}
\usepackage{times,mathptmx}
\usepackage{sectsty}
\usepackage{graphicx} 
\usepackage{lastpage}
\usepackage[format=plain,justification=justified,singlelinecheck=false,font={stretch=1.125,small,sf},labelfont=bf,labelsep=space]{caption}
\usepackage{float}
\usepackage{fancyhdr}
\usepackage{fnpos}
\usepackage[english]{babel}
\usepackage{array}
\usepackage{droidsans}
\usepackage{charter}
\usepackage[T1]{fontenc}
\usepackage[usenames,dvipsnames]{xcolor}
\usepackage{setspace}
\usepackage[compact]{titlesec}
\usepackage{hyperref}

\definecolor{cream}{RGB}{222,217,201}

\begin{document}

\pagestyle{fancy}
\thispagestyle{plain}
\fancypagestyle{plain}{

\fancyhead[C]{\includegraphics[width=18.5cm]{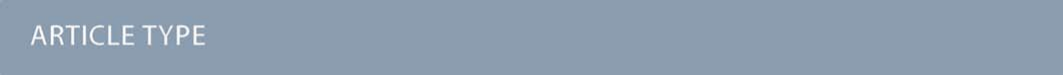}}
\fancyhead[L]{\hspace{0cm}\vspace{1.5cm}\includegraphics[height=30pt]{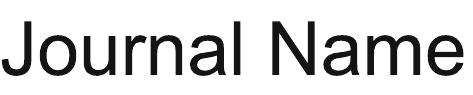}}
\fancyhead[R]{\hspace{0cm}\vspace{1.7cm}\includegraphics[height=55pt]{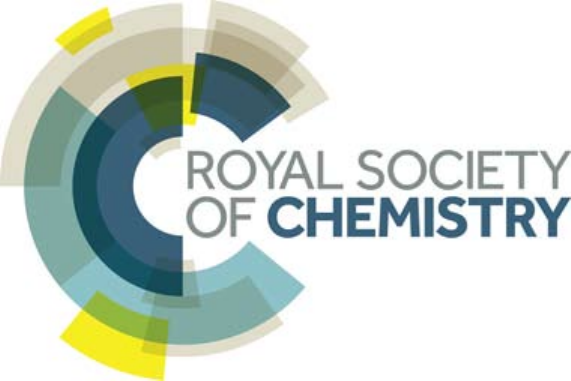}}
\renewcommand{\headrulewidth}{0pt}
}

\makeFNbottom
\makeatletter
\renewcommand\LARGE{\@setfontsize\LARGE{15pt}{17}}
\renewcommand\Large{\@setfontsize\Large{12pt}{14}}
\renewcommand\large{\@setfontsize\large{10pt}{12}}
\renewcommand\footnotesize{\@setfontsize\footnotesize{7pt}{10}}
\makeatother

\renewcommand{\thefootnote}{\fnsymbol{footnote}}
\renewcommand\footnoterule{\vspace*{1pt}%
\color{cream}\hrule width 3.5in height 0.4pt \color{black}\vspace*{5pt}} 
\setcounter{secnumdepth}{5}

\makeatletter 
\renewcommand\@biblabel[1]{#1}            
\renewcommand\@makefntext[1]%
{\noindent\makebox[0pt][r]{\@thefnmark\,}#1}
\makeatother 
\renewcommand{\figurename}{\small{Fig.}~}
\sectionfont{\sffamily\Large}
\subsectionfont{\normalsize}
\subsubsectionfont{\bf}
\setstretch{1.125} 
\setlength{\skip\footins}{0.8cm}
\setlength{\footnotesep}{0.25cm}
\setlength{\jot}{10pt}
\titlespacing*{\section}{0pt}{4pt}{4pt}
\titlespacing*{\subsection}{0pt}{15pt}{1pt}

\fancyfoot{}
\fancyfoot[LO,RE]{\vspace{-7.1pt}\includegraphics[height=9pt]{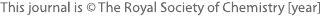}}
\fancyfoot[CO]{\vspace{-7.1pt}\hspace{13.2cm}\includegraphics{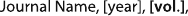}}
\fancyfoot[CE]{\vspace{-7.2pt}\hspace{-14.2cm}\includegraphics{RF}}
\fancyfoot[RO]{\footnotesize{\sffamily{1--\pageref{LastPage} ~\textbar  \hspace{2pt}\thepage}}}
\fancyfoot[LE]{\footnotesize{\sffamily{\thepage~\textbar\hspace{3.45cm} 1--\pageref{LastPage}}}}
\fancyhead{}
\renewcommand{\headrulewidth}{0pt} 
\renewcommand{\footrulewidth}{0pt}
\setlength{\arrayrulewidth}{1pt}
\setlength{\columnsep}{6.5mm}
\setlength\bibsep{1pt}

\makeatletter 
\newlength{\figrulesep} 
\setlength{\figrulesep}{0.5\textfloatsep} 

\newcommand{\topfigrule}{\vspace*{-1pt}%
\noindent{\color{cream}\rule[-\figrulesep]{\columnwidth}{1.5pt}} }

\newcommand{\botfigrule}{\vspace*{-2pt}%
\noindent{\color{cream}\rule[\figrulesep]{\columnwidth}{1.5pt}} }

\newcommand{\dblfigrule}{\vspace*{-1pt}%
\noindent{\color{cream}\rule[-\figrulesep]{\textwidth}{1.5pt}} }

\makeatother

\twocolumn[
  \begin{@twocolumnfalse}
\vspace{3cm}
\sffamily
\begin{tabular}{m{4.5cm} p{13.5cm} }

\includegraphics{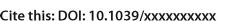} & \noindent\LARGE{\textbf{Theory for polariton-assisted remote energy transfer$^\dag$}} \\
\vspace{0.3cm} & \vspace{0.3cm} \\

 & \noindent\large{Matthew Du,\textit{$^{a}$} Luis A. Mart\'inez-Mart\'inez,\textit{$^{a}$} Raphael F. Ribeiro,\textit{$^{a}$}, Zixuan Hu,\textit{$^{b}$}\textit{$^{c}$} Vinod M. Menon,\textit{$^{d}$}\textit{$^{e}$} and Joel Yuen-Zhou$^{\ast}$\textit{$^{a}$}} \\

\includegraphics{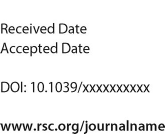} & \noindent\normalsize{Strong-coupling between light and matter produces hybridized states (polaritons) whose delocalization and electromagnetic character allow for novel modifications in spectroscopy and chemical reactivity of molecular systems. Recent experiments have demonstrated remarkable distance-independent long-range energy transfer between molecules strongly coupled to optical microcavity modes. To shed light on the mechanism of this phenomenon, we present the first comprehensive theory of polariton-assisted remote energy transfer (PARET) based on strong-coupling of donor and/or acceptor chromophores to surface plasmons. Application of our theory demonstrates that PARET up to a micron is indeed possible via strong-coupling. In particular, we report two regimes for PARET: in one case, strong-coupling to a single type of chromophore leads to transfer mediated largely by surface plasmons while in the other case, strong-coupling to both types of chromophores creates energy transfer pathways mediated by vibrational relaxation. Importantly, we highlight conditions under which coherence enhances or deteriorates these processes. For instance, while exclusive strong-coupling to donors can enhance transfer to acceptors, the reverse turns out not to be true. However, strong-coupling to acceptors can shift energy levels in a way that transfer from acceptors to donors can occur, thus yielding a chromophore role-reversal or "carnival effect."  This theoretical study demonstrates the potential for confined electromagnetic fields to control and mediate PARET, thus opening doors to the design of remote mesoscale interactions between molecular systems.} \\

\end{tabular}

 \end{@twocolumnfalse} \vspace{0.6cm}

  ]

\renewcommand*\rmdefault{bch}\normalfont\upshape
\rmfamily
\section*{}
\vspace{-1cm}


\footnotetext{\textit{$^{a}$~Department of Chemistry and Biochemistry, University of California San Diego, La Jolla, California 92093, United States. E-mail: jyuenzhou@ucsd.edu}} 
\footnotetext{\textit{$^{b}$~Department of Chemistry, Department of Physics, and Birck Nanotechnology Center, Purdue University, West Lafayette, IN 47907, United States. }}
\footnotetext{\textit{$^{c}$~Qatar Environment and Energy Research Institute, College of Science and Engineering, HBKU, Doha, Qatar. }}
\footnotetext{\textit{$^{d}$~Department of Physics, City College, City University of New York, New York 10031, United States. }}
\footnotetext{\textit{$^{e}$~Department of Physics, Graduate Center, City University of New York, New York 10016, United States. }}

\footnotetext{\dag~Electronic Supplementary Information (ESI) available: [details of any supplementary information available should be included here]. See DOI: 10.1039/b000000x/}



\section{Introduction}

Enhancement of excitation energy transfer (EET) remains an exciting
subfield in the chemical sciences. Although F\"orster resonance energy
transfer (FRET) is one of the most extensively studied and well known
forms of EET, its efficiency range is only at 1-10 nm.\citep{GovorovBookFRET}
As such, exploration of EET schemes beyond that of a traditional pair
of donor and acceptor molecules has been a highly active area of research.
For instance, the theory of multichromophoric FRET\citep{Jang2004}
has been applied to demonstrate the role of coherent exciton delocalization
in photosynthetic light harvesting.\citep{Jang2007,Kassal2013,Lloyd2010,Engel2007,Duque2015}
This coherence, which is due to excitonic coupling between molecular
emitters, has also been argued to increase EET in mesoscopic multichromophoric
assemblies\citep{Scholes2003}, with recent studies even reporting
micron-sized transfer ranges in H-aggregates.\citep{Haedler2015,Saikin2017}
Along similar lines, a well-studied process is plasmon-coupled resonance
energy transfer,\citep{Hsu2017} where molecules which are separated
several tens to hundreds of nanometers apart can efficiently transfer
energy between themselves due to the enhanced electromagnetic fields
provided by the neighboring nanoparticles.\citep{Zhang2014,Ding2017}
Transfer between molecules across even longer distances can be mediated
by the in-plane propagation of surface plasmons (SP), with micron\citep{Bouchet2016,deTorres2016}
and even sub-millimeter\citep{Andrew2004} ranges reported in the
literature. Notably, the plasmonic effects in these last examples
occur in the so-called weak-coupling regime, where the energy exchange
between excitons and plasmons is much slower than their respective
decays.

An intriguing advancement in PARET has recently been reported by the
Ebbesen group for cyanine dye J-aggregates strongly coupled (SC) to
a microcavity mode.\citep{Zhong2017} For spatially separated slabs
of donor and acceptor dyes placed between two mirrors, it was found
that increasing the interslab spacing from 10 to 75 nm led to no change
in the relaxation rate between the hybrid light-matter states or \textit{polaritons},
thus revealing a remarkable distance independence of the process.
Importantly, such PARET phenomenon was already noted in an earlier
work by the Lidzey group\citep{Coles2014} with a different cyanine-dye
system, although the interslab spacing was not systematically varied
there; similar work was also previously reported for hybridization
of Frenkel and Wannier-Mott excitons in an optical microcavity.\citep{Slootsky2014}
Motivated by these experiments, we hereby present a quantum-mechanical
theory for polariton-assisted energy transfer which aims to characterize
the various types of PARET afforded by these hybrid light-matter systems.
To be concrete, we do so within a model where the ``photonic modes''
are SPs in a metal film and consider spatially separated slabs of
donors and acceptor dyes which electrostatically couple to one another
as well as to the SPs. We present a comprehensive formalism which
encompasses the cases where either one or both types of chromophores
are strongly coupled to the SPs. We apply our theory to a model system
similar to those reported by the Ebbesen and Lidzey groups. Our work
complements recent studies proposing schemes to enhance one-dimensional
exciton conductance.\citep{Feist2015,Schachenmayer2015} In those
studies, the delocalization afforded by SC is exploited to overcome
static disorder within the molecular aggregate. Here, the emphasis
is not on disorder (surmountable also by polaritonic topological protection\citep{Yuen-Zhou2016}),
but rather on PARET between two different types of chromophores, where
energy harvested by one chromophore can be collected in another. This
focus on long-range capabilities, as well as in-depth analyses of
the rate contributions for the SC-induced states, provide fresh perspectives
on PARET. In particular, we offer fascinating predictions for the
experimentally unexplored scenario of ``photonic modes'' strongly
coupled to one of donors or acceptors, in which the latter case was
first theoretically investigated for the chromophores in a microcavity.\citep{Basko2000} 

As a preview, we highlight the structure and the main conclusions
of this work (the latter are summarized in Table \ref{tbl:summary}).
We begin by presenting the general Hamiltonian for spatially separated
slabs of donors and acceptor chromophores in Section \ref{sec:Theory}.
EET rates for a single or both chromophores strongly coupled to SPs
are shown in Sections \ref{subsec:case1} and \ref{subsec:case2},
respectively. In the former case, for SC to donors, the rates are
shown to be dependent on spectral overlap and can thus be modified
for either EET enhancement or supression. This result is in stark
contrast with that of strongly coupling acceptors to SPs where, surprisingly,
EET to acceptor polariton states vanishes for large enough samples.
For the case when both chromophores strongly interact with SPs, transfer
is instead mediated by vibrational relaxation, but EET rates are comparable
to the previous case. In Section \ref{sec:application}, we apply
the formalism to study a model system resembling cyanine dye J-aggregates.
Our numerical simulations demonstrate that applying SC to donors only
enables PARET up to 1 micron. We also show that sufficently high SC
to acceptors induces a ``carnival effect'' that reverses the role
of the donor and acceptor. Lastly, when both chromophores are strongly
coupled to SPs, we obtain sizable EET rates at chromophoric separations
over hundreds of nanometers which are in good agreement with experiments.

\begin{table*}
	\caption{Comparison of different cases of PARET arising from SC to donors and/or
		acceptors.}
	\label{tbl:summary} %
	\begin{tabular}{|>{\centering}m{2.5cm}|>{\centering}m{14cm}|}
		\hline 
		SC to & Features\tabularnewline
		\hline 
		Donors only & \begin{itemize}
			\item PARET\emph{ from} donor polariton states; dominated by PRET contribution.
			\item Rate of EET \emph{from} donor dark states $\approx$ bare FRET rate.
		\end{itemize}
		\tabularnewline
		\hline 
		Acceptors only & \begin{itemize}
			\item Low EET \emph{to} acceptor polariton states due to their low density
			of states (compared to dark states) and delocalized character.
			\item Rate of EET \emph{to} acceptor dark states $\approx$ bare FRET rate.
			\item ``Carnival effect'': acceptor and donor can reverse roles.
		\end{itemize}
		\tabularnewline
		\hline 
		Donors and Acceptors & \begin{itemize}
			\item Polariton states hybridize and delocalize donors and acceptors.
			\item Rate of PARET from polariton to dark states $\gg$ rate of PARET from
			dark/polariton to polariton states due to relative density of final
			states. Dark-state manifolds are dense and act as traps.
			\item PARET mediated by vibrational relaxation.
		\end{itemize}
		\tabularnewline
		\hline 
	\end{tabular}
\end{table*}

\section{Theory\label{sec:Theory}}

We begin by describing the polaritonic (plexcitonic) setup that we
theoretically investigate. Let the chromophore slabs lie above ($z>0$)
and parallel to the metal film that sustains SP modes ($z<0$) (example
schematic diagrams are given in Figs. \ref{fgr:dp}a, \ref{fgr:ap}a,
\ref{fgr:reverse}a, and \ref{fgr:pda_d_fixd}a). We assume the metal
film and the slabs are extended along the $xy$ (longitudinal) plane.
The slabs of $C=D,A$ (donor, acceptor) chromophores consist of $N_{xy,C}$,
$N_{z,C}$, and $N_{C}=N_{xy,C}N_{z,C}$ molecules in the $xy$-plane,
$z$-direction, and total, respectively. An effective Hamiltonian
for this setup can be constructed as,
\begin{equation}
	H=H_{D}+H_{A}+H_{P}+H_{DA}+H_{DP}+H_{AP}.\label{eq:gen gen H}
\end{equation}
The term $H_{C}=H_{C}^{(sys)}+H_{C}^{(B)}+H_{C}^{(sys-B)}$ is the
Hamiltonian for the slab with the $C$ chromophores, where (denoting
$\hbar$ as the reduced Planck constant)

\begin{subequations}\label{eq:H_C}
	\begin{align}
		H_{C}^{(sys)} & =\hbar\omega_{C}\sum_{i,j}|C_{ij}\rangle\langle C_{ij}|,\label{eq:x elec}\\
		H_{C}^{(B)} & =\sum_{i,j}\sum_{q}\hbar\omega_{q,C}b_{q,C_{ij}}^{\dagger}b_{q,C_{ij}},\label{eq:xb}\\
		H_{C}^{(sys-B)} & =\sum_{i,j}|C_{ij}\rangle\langle C_{ij}|\sum_{q}\lambda_{q,C}\hbar\omega_{q,C}(b_{q,C_{ij}}^{\dagger}+\text{h.c.}),\label{eq:xeb}
	\end{align}
	
\end{subequations} \noindent represent the system (excitonic), bath
(phononic), and system-bath-coupling contributions, respectively.
The label $C_{ij}$ refers to a $C$ exciton located at the $(i,j)$-th
molecule of the corresponding slab {[}$(i,j)$ indexes an $(xy,z)$
coordinate{]}. We take every $C_{ij}$ exciton to have energy $\hbar\omega_{C}$
and neglect inter-site coupling since it provides an insignificant
contribution to delocalization when compared to the SP couplings.
$b_{q,C_{ij}}^{\dagger}(b_{q,C_{ij}})$ labels the creation (annihilation)
of a phonon of energy $\hbar\omega_{q,C}$ at the $q$-th vibrational
mode of the $(i,j)$-th molecule in the $C$ slab. Given the molecular
character of the problem, vibronic coupling is assumed to be local:
exciton $C_{ij}$ couples linearly to $b_{q,C_{ij}}^{\dagger}$ and
$b_{q,C_{ij}}$ but not to modes in other molecules; these couplings
are characterized by Huang-Rhys factors $\lambda_{q,C}^{2}$. The
SP Hamiltonian $H_{P}=H_{P}^{(sys)}+H_{P}^{(B)}+H_{P}^{(sys-B)}$
has similar form:\citep{WallsBook,Waks2010}

\begin{subequations}\label{eq:H_P}
	\begin{align}
		H_{P}^{(sys)} & =\sum_{\vec{k}}\hbar\omega_{\vec{k}}a_{\vec{k}}^{\dagger}a_{\vec{k}},\label{eq:H_P sys}\\
		H_{P}^{(B)} & =\sum_{q}\hbar\omega_{q,P}b_{q,P}^{\dagger}b_{q,P},\label{eq:H_P B}\\
		H_{P}^{(sys-B)} & =\sum_{\vec{k}}\sum_{q}g_{q,\vec{k}}(b_{q,P}^{\dagger}a_{\vec{k}}+\text{h.c.}),\label{eq:H_P sys-B}
	\end{align}
	
\end{subequations} \noindent where $a_{\vec{k}}^{\dagger}(a_{\vec{k}})$
labels the creation (annihilation) of a SP of energy $\hbar\omega_{\vec{k}}$
and in-plane wavevector $\vec{k}$. Bath modes indexed by $q,P$ with
corresponding operator $b_{q,P}^{\dagger}$ ($b_{q,P}$) and energy
$\hbar\omega_{q,P}$ are coupled to each SP mode $\vec{k}$ with strength
$g_{q,\vec{k}}$. Specifically, these SP interactions occur with either
electromagnetic or phonon modes and represent radiative and Ohmic
losses, respectively.\citep{Waks2010} The remaining rightmost terms
in Eq. (\ref{eq:gen gen H}) represent the dipole-dipole interactions
amongst donors, acceptors, and SP modes. The $H_{DA}$ term is given
by the electrostatic (near-field) dipole-dipole interactions between
donors and acceptors,
\begin{align}
	H_{DA} & =\sum_{i,j}\sum_{l,m}\frac{\mu_{D}\mu_{A}\kappa_{ijlm}}{r_{ijlm}^{3}}(|A_{lm}\rangle\langle D_{ij}|+\text{h.c.}),\label{eq:fret}
\end{align}
where $\mu_{C}=|\vec{\mu}_{C_{ij}}|$ for transition dipole moment
(TDM) $\vec{\mu}_{C_{ij}}$ corresponding to $C_{ij}$, $r_{ijlm}$
is the distance between $D_{ij}$ and $A_{lm}$, and $\kappa_{ijlm}=\hat{\mu}_{D_{ij}}\cdot\hat{\mu}_{A_{lm}}-3(\hat{\mu}_{D_{ij}}\cdot\hat{r}_{ijlm})(\hat{\mu}_{A_{lm}}\cdot\hat{r}_{ijlm})$
is the orientational dependence of the interaction (we have ignored
the corrections to $\kappa_{ijlm}$ due to reflected waves from the
metal\textemdash despite their prominent effects in phenomena such
as photoluminescence\citep{Yuen-Zhou2017}\textemdash since they are
numerically involved\citep{Zhou2011} and do not significantly change
the order of magnitude of the bare dipole-dipole interaction; furthermore
their expected effects in $H_{DA}$ will be overwhelmed by $H_{CP}$,
as we shall explain in Sections \ref{subsec:case1} and \ref{sec:application}).
For simplicity, we take the permittivity on top of the metal to be
a real-valued positive dielectric constant $\epsilon_{d}$. The light-matter
interaction for species $C$ is also dipolar in nature and is described
by\citep{Gonzalez-Tudela2013} 
\begin{align}
	H_{CP} & =\sum_{i,j}\sum_{\vec{k}}\mu_{C}\kappa_{\vec{k}C_{ij}}\sqrt{\frac{\hbar\omega_{\vec{k}}}{2\epsilon_{0}SL_{\vec{k}}}}e^{-a_{d\vec{k}}z_{j,C}}e^{i\vec{k}\cdot\vec{R}_{i,C}}|C_{ij}\rangle\langle G|a_{\vec{k}}+\text{h.c.}\label{eq:sp coupling}
\end{align}
where $(\vec{R}_{i,C},z_{j,C})$ are the position coordinates of $C_{ij}$
and $|G\rangle$ is the electronic ground state (\emph{i.e.}, with
no excitons). Just like in $H_{DA}$, each interaction between an
SP mode and a chromophore (indexed by $\vec{k}$ and $C_{ij}$, respectively)
has an orientation-dependent parameter $\kappa_{\vec{k}C_{ij}}=-\hat{\mu}_{C_{ij}}\cdot(\hat{k}+\frac{k}{a_{d\vec{k}}}\hat{z})$,
where $a_{d\vec{k}}=\sqrt{|\vec{k}|^{2}-\epsilon_{d}(\omega_{\vec{k}}/c)^{2}}$
is the real-valued evanescent SP decay constant on top of the metal.
The light-matter coupling also includes the quantization length $L_{\vec{k}}$
\citep{Archambault2010} and area $S$ of the SP. We refer the reader
to ESI Section \ref{sec:spcoupling} for further details of these
terms.

The general Hamiltonian in Eq. (\ref{eq:gen gen H}) describes a complex
many-body problem consisting of excitons, SPs, and vibrations, all
coupled with each other. To obtain physical insight on the opportunities
afforded by this physical setup, we consider in the next sections
two limit cases where either one or both chromophores are strongly-coupled
to the SP. The study of these two situations already provides considerable
perspective on the wealth of novel EET phenomena hosted by this polaritonic
system. 

\subsection{Case $i$: SC to only one chromophore $C$\label{subsec:case1}}

We consider the case where one of the chromophores $C$ ($D$ or $A$)
is strongly-coupled to an SP but not the other, $C'$. This can happen
when the concentration or thickness of the $C$ slab is sufficiently
high and that of the $C'$ slab low. Under these circumstances, we
write Eq. (\ref{eq:gen gen H}) as $H=H_{0}^{(i)}+V^{(i)}$, where
we define the zeroth-order Hamiltonian as $H_{0}^{(i)}=H_{sys}^{(i)}+H_{B}+H_{sys-B}$.
The system, bath, and their coupling are respectively characterized
by $H_{sys}^{(i)}=H_{D}^{(sys)}+H_{A}^{(sys)}+H_{P}^{(sys)}+H_{CP}$,
$H_{B}=H_{D}^{(B)}+H_{A}^{(B)}+H_{P}^{(B)}$, and $H_{sys-B}=H_{D}^{(sys-B)}+H_{A}^{(sys-B)}+H_{P}^{(sys-B)}$.
The perturbation describing the weak interaction between chromophore
$C'$ and the SC species is $V^{(i)}=H_{DA}+H_{C'P}$. To diagonalize
$H_{sys}^{(i)}$, we introduce a collective exciton basis comprised
of bright $C$ states with in-plane momenta matching those of the
SP modes and ignore the very off-resonant SP-exciton couplings beyond
the first Brillouin zone (FBZ) of the molecular system:\citep{Gonzalez-Tudela2013,Martinez-Martinez2017}
$H_{sys}^{(i)}=H_{C'}+\sum_{\vec{k}\in\text{FBZ}}H_{bright,C}^{(\vec{k})}+H_{dark,C}+\sum_{\vec{k}\notin\text{FBZ}}\hbar\omega_{\vec{k}}a_{\vec{k}}^{\dagger}a_{\vec{k}}$,
where

\begin{subequations}\label{eq:polariton1}
	\begin{align}
		H_{bright,C}^{(\vec{k})} & =\hbar\omega_{C}|C_{\vec{k}}\rangle\langle C_{\vec{k}}|+\hbar\omega_{\vec{k}}a_{\vec{k}}^{\dagger}a_{\vec{k}}+g_{C}(\vec{k})(|C_{\vec{k}}\rangle\langle G|a_{\vec{k}}+\text{h.c.}),\label{eq:hkd}\\
		H_{dark,C} & =H_{C}^{(sys)}-\sum_{\vec{k}\in\text{FBZ}}\hbar\omega_{C}|C_{\vec{k}}\rangle\langle C_{\vec{k}}|.\label{eq:darkkxp}
	\end{align}
	
\end{subequations}\noindent For each $\vec{k}$-mode in the FBZ,
there is only one ``bright'' collective exciton state $|C_{\vec{k}}\rangle=\frac{1}{g_{C}(\vec{k})}\sum_{i,j}\mu_{C}\kappa_{\vec{k}C_{ij}}\sqrt{\frac{\hbar\omega_{\vec{k}}}{2\epsilon_{0}SL_{\vec{k}}}}e^{-a_{d\vec{k}}z_{j,C}}e^{i\vec{k}\cdot\vec{R}_{i,C}}|C_{ij}\rangle$
that couples to the $\vec{k}$-th SP mode $|\vec{k}\rangle=a_{\vec{k}}^{\dagger}|0\rangle$,
where $g_{C}(\vec{k})=\sqrt{\sum_{i,j}\left|\mu_{C}\kappa_{\vec{k}C_{ij}}\sqrt{\frac{\hbar\omega_{\vec{k}}}{2\epsilon_{0}SL_{\vec{k}}}}e^{-a_{d\vec{k}}z_{j,C}}e^{i\vec{k}\cdot\vec{R}_{i,C}}\right|^{2}}$.
In addition to the uncoupled $C'$ states, $H_{sys}^{(i)}$ has two
polariton eigenstates $|\alpha_{C,\vec{k}}\rangle=c_{C_{\vec{k}}\alpha_{C,\vec{k}}}|C_{\vec{k}}\rangle+c_{\vec{k}\alpha_{C,\vec{k}}}|\vec{k}\rangle$
for $\alpha=\text{UP},\text{LP}$ (upper and lower, respectively),
which are also eigenstates of $H_{bright,C}^{(\vec{k})}$ for each
$\vec{k}\in\text{FBZ}$; throughout this work, $c_{mn}=\langle m|n\rangle$.
Furthermore, there is a large reservoir of $N_{C}-N_{xy,C}=N_{xy,C}(N_{z,C}-1)$
``dark'' (purely excitonic) eigenstates $|d_{C,\vec{k}}\rangle$
($d=0,1,\cdots N_{z,C}-2$) which are also eigenstates of $H_{dark,C}$
with bare chromophore energy $\hbar\omega_{C}$. 

EET rates between $C$ and uncoupled $C'$ states can be derived by
applying Fermi's golden rule; the corresponding perturbation $V^{(1)}$
connects vibronic-polariton eigenstates of $H_{0}^{(i)}$ as in FRET
and MC-FRET theories\citep{Jang2002,Jang2003JCP1,Jang2004,Ma2015JCP1}.
For simplicity, we also invoke weak system-bath coupling $H_{sys-B}$,
from which the following expression can be obtained,\citep{Baghbanzadeh2016,Baghbanzadeh2016JPCL}

\begin{equation}
	\gamma_{F\leftarrow I}\approx\frac{2\pi}{\hbar}|\langle F|V^{(i)}|I\rangle|^{2}J_{F,I}.\label{eq:rate general case 1}
\end{equation}
This is the rate of transfer between $H_{sys}^{(i)}$ eigenstates
$|I\rangle$ and $|F\rangle$, where $J_{F,I}$ is the spectral overlap
between absorption and emission spectra, which depend on $H_{B}$
and $H_{sys-B}$ (see Section \ref{subsec:Derivation:-EET-rate} for
derivation of $\gamma_{F\leftarrow I}$ and expression for $J_{F,I}$).
Since our focus is to understand the general timescales expected for
the PARET problem, in Section \ref{sec:application} we treat the
broadening of electronic/polaritonic levels due to $H_{sys-B}$ to
be Lorentzian, although more sophisticated lineshape theories can
be utilized if needed.\citep{MukamelBook} Furthermore, we can in
principle also refine Eq. (\ref{eq:rate general case 1}) to consider
the complexities of vibronic mixing between the various eigenstates
of $H_{sys}^{(i)}$, as done in recent works by Jang and Cao.\citep{Jang2002,Jang2003JCP1,Jang2004,Ma2015JCP1} 

It follows from Eq. (\ref{eq:rate general case 1}) that the rates
from donor states\textemdash either polaritons with given wavevector
$\vec{k}$ or a uniform mixture of dark states with occupation $p_{D_{\vec{k}}}=\frac{1}{N_{D}-N_{xy,D}}$
for all $d,\vec{k}$\textemdash to the incoherent set of all bare
acceptor states are, 

\begin{subequations}\label{eq:rates_to_A}
	\begin{align}
		\gamma_{A\leftarrow\alpha_{D,\vec{k}}} & =\frac{2\pi}{\hbar}\sum_{l,m}|\langle A_{lm}|H_{DA}+H_{AP}|\alpha_{D,\vec{k}}\rangle|^{2}J_{A,\alpha_{D,\vec{k}}},\label{eq:first rate k}\\
		\gamma_{A\leftarrow dark_{D}} & =\frac{2\pi}{\hbar(N_{D}-N_{xy,D})}\sum_{l,m}\sum_{\vec{k}\in\text{FBZ}}\sum_{d}|\langle A_{lm}|H_{DA}|d_{D,\vec{k}}\rangle|^{2}J_{A,d_{D,\vec{k}}},\label{eq:darkD to A}
	\end{align}
	
\end{subequations}\noindent Here, we notice that $\gamma_{A\leftarrow\alpha_{D,\vec{k}}}$
in Eq. (\ref{eq:first rate k}) can be enhanced or supressed relative
to bare (in the absence of metal) FRET due to additional SP-resonance
energy transfer (PRET) channel given by $H_{AP}$, as well as the
spectral overlap $J_{A,\alpha_{D,\vec{k}}}$ that can be modified
by tuning the energy of $|\alpha_{D,\vec{k}}\rangle$. Similar findings
were obtained for electron transfer with only donors strongly coupled
to a cavity mode.\citep{Herrera2016} Given that $|\alpha_{D,\vec{k}}\rangle$
corresponds to a delocalized state, one would expect a superradiant
enhancement of the rate;\citep{Lloyd2010,Kassal2013} in practice,
this effect is minor due to the distance dependence of $H_{DA}$ (see
Section \ref{subsec:no superradiant enhancement}). On the other hand,
Eq. (\ref{eq:darkD to A}) presents an averaged rate $\gamma_{A\leftarrow dark_{D}}$
from the dark states and hence does not feature a PRET term. In fact,
it converges (see Section \ref{subsec:derivation_dark donors} for
derivation) to the bare FRET rate (Eq. (\ref{eq:darkD to A-1}) below)
in the limit of large $N_{D}\gg N_{xy,D}$ (when there are many layers
of chromophores along $z$) and isotropically averaged and orientationally
uncorrelated TDMs $\vec{\mu}_{C_{ij}}$ for both $C=D,A$.

In contrast, strongly coupling the acceptor states to SPs yields the
following rates:

\begin{subequations}\label{eq:rates_to_D}
	\begin{align}
		\gamma_{\alpha_{A}\leftarrow D} & =\frac{2\pi}{\hbar N_{D}}\sum_{\vec{k}\in\text{FBZ}}\sum_{i,j}|\langle\alpha_{A,\vec{k}}|H_{DA}+H_{DP}|D_{ij}\rangle|^{2}J_{\alpha_{A,\vec{k}},D},\label{eq:second rate k}\\
		\gamma_{dark_{A}\leftarrow D} & =\frac{2\pi}{\hbar N_{D}}\sum_{\vec{k}\in\text{FBZ}}\sum_{d}\sum_{i,j}|\langle d_{A,\vec{k}}|H_{DA}|D_{ij}\rangle|^{2}J_{d_{A,\vec{k}},D}.\label{eq:last rate 1}
	\end{align}
	
\end{subequations}\noindent Here, we have calculated average rates
over the $N_{D}$ possible initial states at the $D$ slab and summed
over all final states for each polariton/dark band. Given the asymmetry
of Fermi's golden rule with respect to initial and final states (rates
scale with the \emph{probabilities} of occupation of \emph{initial}
states and with the \emph{density} of \emph{final} states),\citep{Kassal2013}
the physical consequences of Eq. (\ref{eq:rates_to_D}) are quite
different to those of its counterpart in Eq. (\ref{eq:rates_to_A})
when $N_{A}\gg N_{xy,A}$ and all TDMs are isotropically averaged
and feature no orientational correlations amongst them. First, it
is interesting to note that the rate of EET to the polariton states
is reduced substantially compared to the bare FRET rate when the average
$|\langle A_{lm}|H_{DA}|D_{ij}\rangle|^{2}$ is greater than the average
$|\langle\vec{k}|H_{DP}|D_{ij}\rangle|^{2}$ (see Section \ref{subsec:derivations sc acceptors}
for a formal derivation of this statement). At first sight, this appears
to be counterintuitive in light of the various recently reported phenomena
which are enhanced upon exciton delocalization.\citep{Scholes2017}
Nevertheless, this statement is actually easily understood from a
final-density-of-states argument (see Eq. (\ref{eq:second rate k})):
$\sum_{\vec{k}\in\text{FBZ}}$ reflects the $N_{xy,A}$ bright acceptor
collective modes that contrbute to $\gamma_{\alpha_{A}\leftarrow D}$,
as opposed to the $N_{A}$ localized acceptor states that contribute
to the bare FRET rate. On the other hand, $\gamma_{dark_{A}\leftarrow D}$
behaves similarly to Eq. (\ref{eq:darkD to A}) in that it converges
(see Eqs. (\ref{eq:last rate 1-1}) and (\ref{eq:d to pa_expl}))
to the bare FRET rate. Thus, at donor-acceptor separations where the
square of the coupling for FRET exceeds on average that for PRET,
the inequality $\gamma_{\alpha_{A}\leftarrow D}\ll\gamma_{dark_{A}\leftarrow D}$
is expected to hold. 

Our analyses of Eqs. (\ref{eq:rates_to_A}) and (\ref{eq:rates_to_D})
reveal one of the main conclusions of this letter: \emph{while strongly
	coupling to $D$ but not to $A$ might yield a $D\to A$ EET rate
	change with respect to the bare case, strong coupling to $A$ but
	not to $D$ will change that process in a negligible manner}. Interestingly,
these trends have also been observed for transfer between layers of
donor and acceptor quantum dots selectively coupled to metal nanoparticles
in the weak-interaction regime.\citep{Ozel2013} However, polariton
formation with $A$ is not useless, for one may consider the interesting
prospect of converting $A$ states into new donors. As we shall show
in the next paragraphs, this role reversal or ``carnival effect''
can be achieved when the UP is higher in energy than the bare donor
states. These findings are quite general and should apply to other
molecular processes as long as the interactions between reactants
and products (taking the roles of donors and acceptors) also decay
at large distances, a scenario that is chemically ubiquitous.\citep{Hutchison2012}

\subsection{Case $ii$: SC to both donor and acceptor chromophores\label{subsec:case2}}

We next consider strongly coupling SPs to both donors and acceptors.
We rewrite Eq. (\ref{eq:gen gen H}) as $H=H_{0}^{(ii)}+V^{(ii)}$,
where $H_{0}^{(ii)}=H_{sys}^{(ii)}+H_{B}$ and the perturbation is
$V^{(ii)}=H_{sys-B}$. Here, $H_{sys}^{(ii)}=H_{D}^{(sys)}+H_{A}^{(sys)}+H_{DP}+H_{AP}+H_{DA}+H_{P}$
is the polariton Hamiltonian. The EET pathways of interest become
those where $H_{sys-B}$ induces vibrationally mediated relaxation
\textit{among} the delocalized states resulting from SC. The transfer
rates describing these processes can be deduced by Fermi's golden
rule too, the resulting expressions coinciding with those derived
with Redfield theory.\citep{MayBook} Although not necessary, we take
$N_{xy,D}=N_{xy,A}=N_{xy}$ to avoid mathematical technicalities about
working with two $\vec{k}$ grids of different sizes, a complication
that does not give more insight into the physics of interest. As done
in Section \ref{subsec:case1}, we rewrite $H_{sys}^{(ii)}$ in $\vec{k}$-space:
$H_{sys}^{(ii)}=\sum_{\vec{k}\in\text{FBZ}}H_{bright}^{(\vec{k})}+H_{dark,D}+H_{dark,A}+H_{DA}+\sum_{\vec{k}\notin\text{FBZ}}\hbar\omega_{\vec{k}}a_{\vec{k}}^{\dagger}a_{\vec{k}}$,
where
\begin{align}
	H_{bright}^{(\vec{k})} & =\hbar\omega_{D}|D_{\vec{k}}\rangle\langle D_{\vec{k}}|+\hbar\omega_{A}|A_{\vec{k}}\rangle\langle A_{\vec{k}}|+\hbar\omega_{\vec{k}}a_{\vec{k}}^{\dagger}a_{\vec{k}}\label{eq:brightdak}\\
	& \qquad+g_{D}(\vec{k})(|D_{\vec{k}}\rangle\langle G|a_{\vec{k}}+\text{h.c.})+g_{A}(\vec{k})(|A_{\vec{k}}\rangle\langle G|a_{\vec{k}}+\text{h.c.}),\nonumber 
\end{align}
and the terms labeled $dark$ are defined analogously to those in
Eq. (\ref{eq:darkkxp}). For each $\vec{k}\in\text{FBZ}$, there are
three polariton eigenstates of $H_{bright}^{(\vec{k})}$ that are
linear combinations of $|D_{\vec{k}}\rangle$, $|A_{\vec{k}}\rangle$,
and $|\vec{k}\rangle$, and we call them UP, middle polariton (MP),
and LP, according to their energy ordering. In addition, the presence
of $H_{dark,C}$ yields $N_{C}-N_{xy,C}$ dark $C$ eigenstates. 

The resulting expressions for the rates of transfer from a single
polariton state or average dark state to an entire polariton or dark
state bands are

\begin{subequations}\label{eq:rates_both}
	\begin{align}
		\gamma_{\beta\leftarrow\alpha_{\vec{k}}} & =\sum_{\vec{k}'\in\text{FBZ}}\sum_{C}|c_{C_{\vec{k}'}\beta_{\vec{k}'}}|^{2}|c_{C_{\vec{k}}\alpha_{\vec{k}}}|^{2}\sum_{i,j}|c_{C_{ij}C_{\vec{k}'}}|^{2}|c_{C_{ij}C_{\vec{k}}}|^{2}\mathcal{R}_{C}(\omega_{\beta_{\vec{k}'}\alpha_{\vec{k}}}),\label{eq:bright to bright}\\
		\gamma_{\alpha\leftarrow dark_{C}} & =\frac{1}{N_{C}-N_{xy}}\sum_{\vec{k}'\in\text{FBZ}}\sum_{\vec{k}\in\text{FBZ}}\sum_{d}|c_{C_{\vec{k}'}\alpha_{\vec{k}'}}|^{2}|c_{C_{\vec{k}}d_{C,\vec{k}}}|^{2}\nonumber \\
		& \times\sum_{i,j}|c_{C_{ij}C_{\vec{k}'}}|^{2}|c_{C_{ij}C_{\vec{k}}}|^{2}\mathcal{R}_{C}(\omega_{\alpha_{\vec{k}'}C}),\label{eq:from dark}\\
		\gamma_{dark_{C}\leftarrow\alpha_{\vec{k}}} & =\sum_{\vec{k}'\in\text{FBZ}}\sum_{d}|c_{C_{\vec{k}'}d_{C,\vec{k}'}}|^{2}|c_{C_{\vec{k}}\alpha_{\vec{k}}}|^{2}\sum_{i,j}|c_{C_{ij}C_{\vec{k}'}}|^{2}|c_{C_{ij}C_{\vec{k}}}|^{2}\mathcal{R}_{C}(\omega_{C\alpha_{\vec{k}}}),\label{eq:last rate k}
	\end{align}
	
\end{subequations}\noindent for $\alpha,\beta=\text{UP},\text{MP},\text{LP}$
and $C=D,A$ (see Section \ref{subsec:derivations_bb} for derivation
of Eqs. (\ref{eq:rates_both})). To intuitively understand Eq. (\ref{eq:bright to bright}),
note that $|c_{C_{\vec{k}}\alpha_{\vec{k}}}|^{2}|c_{C_{ij}C_{\vec{k}}}|^{2}$
and $|c_{C_{\vec{k}'}\beta_{\vec{k}'}}|^{2}|c_{C_{ij}C_{\vec{k}'}}|^{2}$
are the fractions of exciton $|C_{ij}\rangle$ in the polariton states
$|\alpha_{\vec{k}}\rangle$ and $|\beta_{\vec{k}'}\rangle$, respectively,
while $\mathcal{R}_{C}(\omega_{\beta_{\vec{k}'}\alpha_{\vec{k}}})$
is the single-molecule rate of vibrational relaxation at the energy
difference $\omega_{\beta_{\vec{k}'}}-\omega_{\alpha_{\vec{k}}}$.
More specifically, $\mathcal{\mathcal{R}}_{C}(\omega)=2\pi\Theta(-\omega)[n(-\omega)+1]\mathcal{J}_{C}(-\omega)+2\pi\Theta(\omega)n(\omega)\mathcal{J}_{C}(\omega)$,
where $\Theta(\omega)$ is the Heaviside step function, $n(\omega)=\frac{1}{e^{\hbar\omega/k_{B}T}-1}$
is the Bose-Einstein distribution function ($k_{B}$ is the Boltzmann
constant and $T$ is temperature) for zero chemical potential $\mu=0$,
and $\mathcal{J}_{C}(\omega)=\sum_{q}\lambda_{q,C}^{2}\omega_{q,C}^{2}\delta(\omega-\omega_{q,C})$
is the spectral density for chromophore $C$.\citep{MayBook} Hence,
one can interpret $\beta\leftarrow\alpha_{\vec{k}}$ as a sum of incoherent
processes (over $C$ and $i,j$) where the (local) vibrational modes
in $C_{ij}$ absorb or emit phonons concomittantly inducing population
transfer between the various eigenstates of $H_{sys}^{(ii)}$. Eqs.
(\ref{eq:from dark}) and (\ref{eq:last rate k}) can be interpreted
in a similar light. We shall comment on some important qualitative
trends in these rates while for simplicity assuming that $N_{z,D}=N_{z,A}=N_{z}$.
First, EET from polariton or dark states to a polariton band (Eqs.
(\ref{eq:bright to bright}) and (\ref{eq:from dark})) scale as $\frac{\mathcal{R}_{C}}{N_{z}}$.
To see this, note that both $|c_{C_{\vec{k}'}\beta_{\vec{k}'}}|^{2}$
and $|c_{C_{\vec{k}}\alpha_{\vec{k}}}|^{2}$ are $O(1)$, while $|c_{C_{ij}C_{\vec{k}'}}|^{2}$
and $|c_{C_{ij}C_{\vec{k}}}|^{2}$ are $O(\frac{1}{N_{xy}N_{z}})$,
but the summations $\sum_{\vec{k}\in\text{FBZ}}$ and$\sum_{i,j}$
are respectively carried over $N_{xy}$ and $N_{xy}N_{z}$ terms.
On the other hand, $\gamma_{dark_{C}\leftarrow\alpha_{\vec{k}}}$
takes values that are on the order of the single-molecule decay $\mathcal{R}_{C}(\omega_{C\alpha_{\vec{k}}})$.
For sufficiently large $N_{z}$, these scalings are consistent with
previous studies on relaxation dynamics of polaritons\citep{Agranovich2003,Litinskaya2004,delPino2015}
and can be summarized as follows: the dominant channels of relaxation
are from the polariton states to a reservoir of dark states that share
the same exciton character; their timescales are comparable to those
of the corresponding single-chromophore vibrational relaxation; given
the large density of states in this reservoir compared to the polariton
bands, the dark states act as a population sink or trap from which
population can only leak out very slowly.\citep{Canaguier-Durand2015,delPino2015} 

\section{Application of the theory\label{sec:application}}

The theory above is now applied to study EET kinetics associated with
slabs of chromophores with $\hbar\omega_{D}=2.1\text{ eV}$, $\hbar\omega_{A}=1.88\text{ eV}$;
these transition energies are chosen to match those of the J-aggregated
cyanine dyes (TDBC and BRK5714, respectively) used in previous polariton
experiments \citep{Zhong2016,Zhong2017}. For simplicity, this section
assumes $T=0$ and thus only considers downhill transfers to/from
polariton and dark states. We describe the metal with Drude permittivity
of silver ($\omega_{p}=9.0\text{ eV}$, $\epsilon_{\infty}=1$;\citep{PalikBook}
see Section \ref{sec:spcoupling}) and all medium at $z>0$ (including
molecular slabs) with $\epsilon_{d}=1$. We model spectral overlaps
(Eqs. (\ref{eq:rates_to_A}) and (\ref{eq:rates_to_D})) with Lorentzian
functions $J_{F,I}=\frac{\frac{\Gamma_{I}+\Gamma_{F}}{2}}{\pi[(\frac{\Gamma_{I}+\Gamma_{F}}{2})^{2}+(\hbar\omega_{FI})^{2}]}$
whose parameters are estimated as in Section \ref{subsec:lorentzian };
we set $\Gamma_{A}\approx\Gamma_{D}=47\text{ meV}$ to represent observed
values for absorption of TDBC\citep{Bellessa2004} and $\Gamma_{P,\vec{k}}=\frac{v_{g}(\vec{k})}{L_{\vec{k}}}$,\citep{Gonzalez-Tudela2013}
where $v_{g}(\vec{k})$ is the SP group velocity. Rigorous treatments
of lineshape functions have been previously reported in MC-FRET literature
and could be applied to this problem as well,\citep{Jang2003JCP1,Ma2015JCP1,Ma2015JCP2,Moix2015}
although this effort is beyond the scope of our work. We even neglect
differences in TDMs and assign $\mu_{D}=\mu_{A}=10\text{ D}$, a typical
number for cyanine-dyes.\citep{Valleau2012} 

\begin{figure}
	\centering\includegraphics{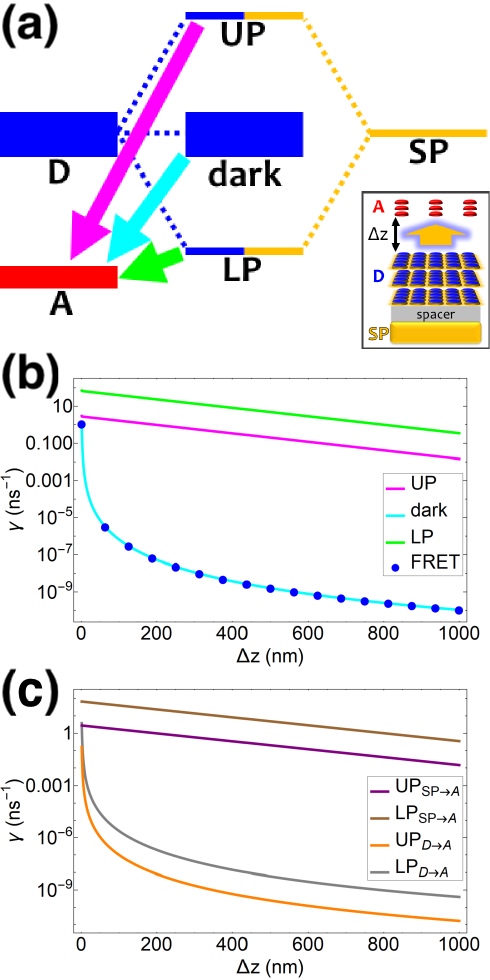}
	
	\caption{(a) Schematic energy-level diagram showing the EET transitions from
		donors strongly coupled to SPs to bare acceptors . The thickness of
		the horizonal lines denotes the density of states while the thickness
		of arrows corresponds to rate of transition (thicknesses not drawn
		to scale). Inset: representation of the EET process from a thick and
		dense slab of donors (featuring SC to SPs) to a dilute monolayer of
		acceptors. (b) Rates as a function of donor-acceptor separation $\Delta z$
		for EET from donor polariton and dark states to acceptors (lines).
		The rate from dark states and for the bare-donor FRET (dots), are
		calculated in the same manner. (c) Contributions of rates for transfer
		from donor UP and LP to acceptor states due to donor-acceptor and
		SP-acceptor interactions.}
	\label{fgr:dp} 
\end{figure}

We now proceed to simulations for Case 1, where only one of the molecular
species forms polaritons. For simplicity, we assume isotropically
oriented and spatially uncorrelated dipoles, upon which we find the
interesting observation that the transfer rates in Eq. (\ref{eq:rates_to_A})
can be approximately decomposed into incoherent sums of FRET and PRET
rates (see Sections \ref{subsec:derivation_donor polaritons} and
\ref{subsec:derivation_dark donors} for more explicit expressions,
derivations, and justification of validity),

\begin{subequations}\label{eq:rates_to_A_uncorr}
	
	\begin{align}
		\gamma_{A\leftarrow\alpha_{D,\vec{k}}} & \approx\frac{2\pi}{\hbar}\sum_{l}\sum_{i,j}\left(|c_{D_{\vec{k}}\alpha_{D,\vec{k}}}|^{2}|c_{D_{ij}D_{\vec{k}}}|^{2}|\langle A_{l0}|H_{DA}|D_{ij}\rangle|^{2}\right.\nonumber \\
		& \left.+|c_{\vec{k}\alpha_{D,\vec{k}}}|^{2}|\langle A_{l0}|H_{AP}|\vec{k}\rangle|^{2}\right)J_{A,\alpha_{D,\vec{k}}},\label{eq:first rate k-1}\\
		\gamma_{A\leftarrow dark_{D}} & =\gamma_{\text{bare FRET}}=\frac{2\pi}{\hbar N_{D}}\sum_{l}\sum_{i,j}|\langle A_{l0}|H_{DA}|D_{ij}\rangle|^{2}J_{A,D},\label{eq:darkD to A-1}
	\end{align}
	
\end{subequations}\noindent where, as explained above, only $\gamma_{A\leftarrow\alpha_{D,\vec{k}}}$
differs from $\gamma_{\text{bare FRET}}$. More concretely, we consider
a 35-nm-thick slab of donors with $1\times10^{9}\text{ }\text{molecules}/\mu\text{m}^{3}$
on top of a 1 nm spacer placed on a plasmonic metal film. We set the
monolayer slab of acceptors with $1\times10^{4}\text{ }\text{molecules}/\mu\text{m}^{2}$
at varying distances from the donors (Fig. \ref{fgr:dp}a). Then the
collective couplings of the donor-resonant SP mode at $|\vec{k}|=1.1\times10^{7}\text{ m}^{-1}$
to donors and acceptors is $g_{D}=155\text{ meV}$ and $g_{A}\leq2.5\text{ meV}$,
respectively. When there is no separation between donor and acceptor
slabs, rates $>1\text{ ns}^{-1}$ (Fig. \ref{fgr:dp}b) are obtained
for transfer to acceptors from the UP ($\sim10\text{ ns}^{-1})$,
LP ($\sim100\text{ ns}^{-1})$, or the set of dark states ($\sim10\text{ ns}^{-1}$).
As separation increases however, the rate from dark states decays
much faster than those from either UP or LP. This difference stems
from the slowly decaying PRET contribution of the polaritons, as well
as the totally excitonic character of the dark states, which can only
undergo FRET but not PRET (Fig. \ref{fgr:dp}a,b). In fact, for large
distances, the FRET contribution becomes significantly overwhelmed
by PRET (Fig. \ref{fgr:dp}c), in consistency with previous studies
in the weak SP-coupling regime.\citep{Govorov2007} As the distance
between slabs approaches $1\text{ }\mu\text{m}$, it is fascinating
that while transfer from dark states (and thus bare FRET) practically
vanishes, the rate from either UP ($\sim1\text{ ns}^{-1}$) or LP
($\sim0.01\text{ ns}^{-1}$) is still on the order of typical fluorescence
lifetimes.\citep{ValeurBook} In FRET language, this PARET can be
said to have a Förster distance in the $\mu\text{m}$ range, or to
be 1000-fold greater than the typical nm-range.\citep{MedintzBook}
Interestingly, the LP rate exceeds the UP one by 1-2 orders of magnitude
at all separations due to greater spectral overlap with the acceptor
(Figs. \ref{fgr:dp}a and \ref{fgr:pd_overlaps}). 

\begin{figure}
	\centering\includegraphics{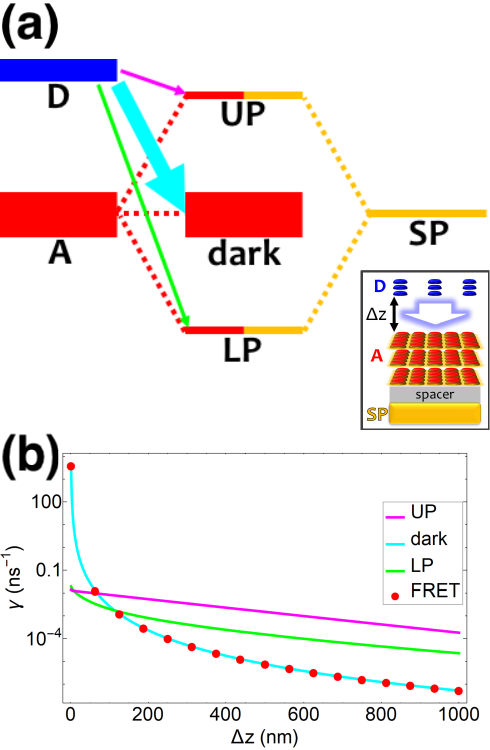}
	
	\caption{(a) Schematic energy-level diagram showing the EET transitions from
		bare donors to acceptors strongly coupled to SPs. The thickness of
		the horizonal lines denotes the density of states while the thickness
		of arrows corresponds to rate of transition (thicknesses not drawn
		to scale). Inset: representation of the EET process from a dilute
		monolayer of donors to a thick and dense slab of acceptors (featuring
		SC to SPs). (b) Rates as a function of donor-acceptor separation $\Delta z$
		for energy transfer from donors to acceptor polaritons and dark states
		(lines). The rate to dark states and for bare-acceptor FRET (dots),
		are calculated in the same manner. }
	\label{fgr:ap} 
\end{figure}

In contrast, strongly coupling the acceptors to a resonant SP mode
does not lead to the aforementioned PARET from donors to acceptors
(Fig. \ref{fgr:ap}). Making the same assumptions as above about the
isotropically oriented and spatially uncorrelated dipoles gives (Section
\ref{subsec:derivations sc acceptors}),

\begin{subequations}\label{eq:rates_to_D_uncorr}
	
	\begin{align}
		\gamma_{\alpha_{A}\leftarrow D} & \approx\frac{2\pi}{\hbar N_{D}}\sum_{\vec{k}\in\text{FBZ}}\sum_{i}\left(\sum_{l,m}|c_{A_{\vec{k}}\alpha_{A,\vec{k}}}|^{2}|c_{A_{lm}A_{\vec{k}}}|^{2}|\langle A_{lm}|H_{DA}|D_{i0}\rangle|^{2}\right.\nonumber \\
		& \left.+|c_{\vec{k}\alpha_{A,\vec{k}}}|^{2}|\langle\vec{k}|H_{DP}|D_{i0}\rangle|^{2}\right)J_{\alpha_{A,\vec{k}},D},\label{eq:second rate k-1}\\
		\gamma_{dark_{A}\leftarrow D} & =\gamma_{\text{bare FRET}}'=\frac{2\pi}{\hbar N_{D}}\sum_{i}\sum_{l,m}|\langle A_{lm}|H_{DA}|D_{i0}\rangle|^{2}J_{A,D}.\label{eq:last rate 1-1}
	\end{align}
	
\end{subequations}\noindent We consider (Fig. \ref{fgr:ap}a) a 50
nm-thick acceptor slab with a concentration of $1\times10^{9}\text{ \text{molecules}}/\mu\text{m}^{3}$
on top of the 1 nm spacer placed on the metal, and a monolayer of
donors with concentration $1\times10^{4}\text{ }\text{molecules}/\mu\text{m}^{2}$
at varying distances from the acceptors. Notice that $\gamma_{dark_{A}\leftarrow D}$
becomes another bare FRET rate like in Eq. (\ref{eq:darkD to A-1}).
For $\gamma_{\alpha_{A}\leftarrow D}$, we still see that PRET still
dominates over FRET for long distances (Fig. \ref{fgr:pa_breakdown}).
However, due to the suppression of $\gamma_{\alpha_{A}\leftarrow D}$
relative to $\gamma_{dark_{A}\leftarrow D}$ explained in Section
\ref{subsec:case1}, the limited spatial range of interactions of
$H_{DA}$ and $H_{DP}$, and the fact that the donor energy is lower
than that of $|\text{UP}_{A,\vec{k}}\rangle$for most $\vec{k}\in\text{FBZ}$
(Fig. \ref{fgr:pa_disp}), the rates to acceptor polaritons fall below
fluorescence timescales\citep{ValeurBook} and therefore offer no
meaningful enhancements with respect to the bare FRET case (Fig. \ref{fgr:ap}b). 

\begin{figure}
	\centering\includegraphics{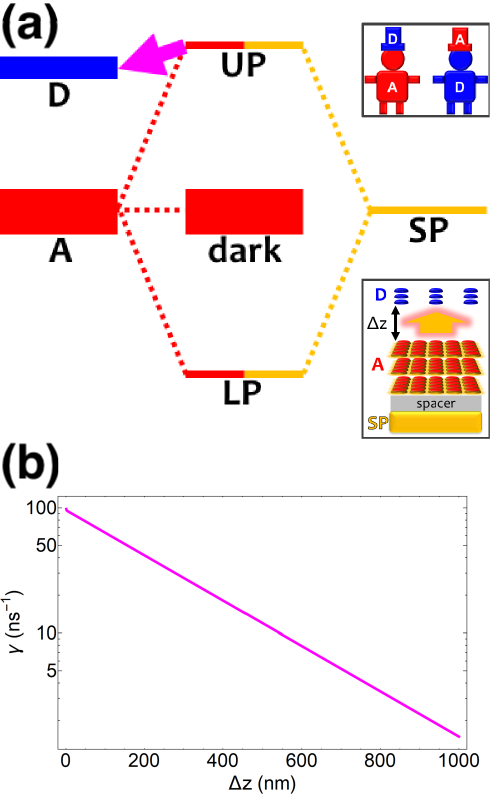}
	
	\caption{(a) Schematic energy-level diagram showing the ``carnival-effect''-EET
		role-reversal process from acceptor UP state to bare donors. Insets:
		(top) cartoon illustrating the ``carnival effect'' between donors
		and acceptors and (bottom) representation of the reversed-role EET
		process from a thick and dense slab of acceptors (featuring SC to
		SPs) to a dilute monolayer of donors. (b) Rate as a function of donor-acceptor
		separation $\Delta z$ for energy transfer from acceptor UP to bare
		donors. }
	\label{fgr:reverse} 
\end{figure}

Coupling SPs to acceptors need not, however, be a disapointment. Increasing
the collective coupling to $g_{A}=237\text{ meV}$ while keeping $g_{D}=1.7\,\text{meV}$
lifts the acceptor UP energy $\hbar\omega_{UP_{A,\vec{k}}}$ to be
higher than $\hbar\omega_{D}$ (Figs. \ref{fgr:reverse}a and \ref{fgr:pd_reverse_overlaps}),
thus allowing for the carnival effect where the donors and acceptors
reverse roles. Due to sufficient spectral overlap between the acceptor
UP and donor states (Figs. \ref{fgr:reverse}a and \ref{fgr:pd_reverse_overlaps}),
transfer \textit{from} UP occurs at $\sim100\text{ ns}^{-1}$ for
donor-acceptor separation of 1 nm and drops only to $\sim1\text{ ns}^{-1}$
when this separation approaches $1\text{ }\mu\text{m}$ (Fig. \ref{fgr:reverse}b).
On the other hand, neither the acceptor dark nor LP states contribute
to this reversed PARET given their lack of spectral overlap with the
donors and detailed balance (especially at $T=0$), . This result
provides the second main conclusion of our work: \textit{polaritons
	offer great versatility to control spectral overlaps without actual
	chemical modifications to the molecules and can therefore endow them
	with new physical properties}. Before proceeding to simulations for
Case $ii$, it should first be noted that while our model neglects
intermediate- and far-field donor-acceptor dipole-dipole interactions
that become relevant at $\sim\mu\text{m}$ distances, these couplings
are expected to be small compared to PRET couplings and therefore
should not change our results for Case $i$ qualitatively but can
be modeled according to previous literature.\citep{Dung2002-2} Second,
we highlight that the PARET from UP to donors for the donors-only
and reversed cases of SC may not be readily observable in experiments
due to their competition with fast vibrational relaxation to dark
states (\textasciitilde{}10-100 fs for exciton-microcavity systems),\citep{Coles2011,Virgili2011,Coles2011PRB,Somaschi2011,Agranovich2003}
as will be discussed next for both donors and acceptors strongly coupled
to SPs.

\begin{figure}
	\centering\includegraphics{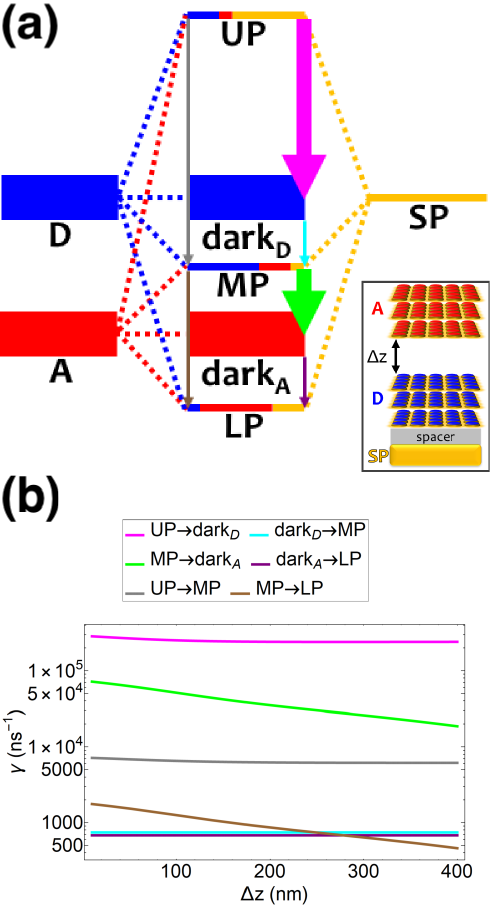}
	
	\caption{(a) Schematic energy-level diagram showing the EET transitions among
		polariton and dark states for both donors and acceptors strongly coupled
		to SPs. The SP mode is resonant with the donor transition; the donor
		slab lies 1 nm above the metal and has fixed position ($z>0$) while
		the acceptor slab is moved in the $+z$-direction to vary $\Delta z$.
		The thickness of the horizonal lines denotes the density of states
		while the thickness of arrows corresponds to rate of transition (thicknesses
		not drawn to scale). Inset: representation of the setup featuring
		thick and dense slabs for both types of chromophores. (b) Rates for
		selected downhill transitions as a function of donor-acceptor separation
		$\Delta z$.}
	\label{fgr:pda_d_fixd} 
\end{figure}

Finally, for strongly coupling both chromophores to the same SP mode,
we limit ourselves to donor-acceptor separations $\geq10$ nm to ignore
$H_{DA}$ terms (an approximation validated by the calculations above
demonstrating that at such distances, rate contributions of PRET overwhelm
those of FRET ). The thickness (35 nm) and density ($1\times10^{9}\text{ }\text{molecules}/\mu\text{m}^{-3}$)
of each slab is large enough to allow for SC of a SP mode to both
chromophores separated by $\sim400$ nm (Fig. \ref{fgr:pda_d_fixd_coupling}),
even though we set the donor to be in resonance with the SP (Fig.
\ref{fgr:pda_d_fixd}). To evaluate the rates derived (Section \ref{subsec:rates_both})
from Eq. (\ref{eq:rates_both}) under condition $N_{C}\gg N_{xy,C}$
for $C=D,A$, we introduce a spectral density representing intramolecular
exciton-phonon coupling of TDBC: $\mathcal{J}_{A}(\omega)\approx\mathcal{J}_{D}(\omega)=\mathcal{J}(\omega)$,
where
\begin{equation}
	\mathcal{J}(\omega)=\sum_{q\in B_{\text{TDBC}}}\lambda_{q}^{2}\omega_{q}^{2}\frac{\frac{\Gamma/\hbar}{2}}{\pi[(\frac{\Gamma/\hbar}{2})^{2}+(\omega-\omega_{q})^{2}]},\label{eq:spectral density tdbc}
\end{equation}
$\Gamma=47\text{ meV}$ is equal to the chromophore decay energy,
and $B_{\text{TDBC}}$ is the discrete set (Section \ref{subsec:rates_both})
of localized vibrational modes which significantly couple to each
TDBC exciton; such coupling has been experimentally\citep{Coles2011,Virgili2011,Coles2011PRB,Somaschi2011,Coles2013}
and theoretically supported as the mechanism of vibrational relaxation
for the dye\citep{Agranovich2003,Litinskaya2004,Chovan2008,Michetti2008,Michetti2009};
our spectral density has been reconstructed from the works of Agranovich
and coworkers \citep{Agranovich2003,Litinskaya2004}. By placing the
donor slab on top of the spacer on the metal and varying the acceptor
position on top of the donors (Fig. \ref{fgr:pda_d_fixd}a), we find
that for all donor-acceptor separations, the rates of PARET from UP
to dark donors ($\sim10^{5}\text{ ns}^{-1}$) and MP to dark acceptors
($\sim10^{4}-10^{5}\text{ ns}^{-1}$) are substantially higher compared
to those from dark donors to MP ($\sim10^{3}\text{ ns}^{-1}$) and
dark acceptors to LP ($\sim10^{3}\text{ ns}^{-1}$) (Fig. \ref{fgr:pda_d_fixd}b).
These observations are in agreement with our discussion above, where
the dark state manifolds act as population sinks due to their high
density of states. Indeed, we also notice that the rates for $\text{UP}\rightarrow dark_{D}$
and $\text{MP}\rightarrow dark_{A}$ are enhanced (Fig. \ref{fgr:pda_d_fixd}b)
compared to those of $\text{UP}\rightarrow\text{MP}$ and $\text{MP}\rightarrow\text{LP}$,
respectively, by approximately $N_{z}=35$, the analytically estimated
ratio solely based on the associated density of final states. Another
interesting detail seen from Fig. \ref{fgr:pda_d_fixd}b is that the
$\text{UP}\rightarrow\text{MP}$ ($\text{MP}\rightarrow\text{LP}$)
rate with respect to the interslab distance $\Delta z$ is essentially
parallel to that of $\text{UP}\rightarrow dark_{D}$ ($\text{MP}\rightarrow dark_{A}$).
This is a consequence of the MP (LP) having mostly donor (acceptor)
energy (Fig. \ref{fgr:pda_d_fixd_disp}) and character (Fig. \ref{fgr:pda_d_fixd_char})
for most points in the FBZ. The $\text{UP}\rightarrow dark_{A}$ and
$\text{UP}\rightarrow\text{LP}$ processes behave similarly (Fig \ref{fgr:pda_d_fixd_minor}). 

These calculated rates even establish consistency with a number of
recent notable experiments. First, our results corroborate the experimental
observation of efficient PARET for separated donor and acceptor slabs
of cyanine dyes strongly coupled to a microcavity.\citep{Zhong2017,Coles2014}
While our SP model for the SC of both excitons cannot account for
the exact distance-independent PARET\citep{Zhong2017} amongst donor
and acceptor slabs in a microcavity, the rates are essentially constant
over hundreds of nanometers due to the slowly decaying SP fields.
Additional validation of our theory can be obtained by comparing directly
to experimentally fitted rates for a blend of two cyanine dyes where
physical separation of the dyes did not significantly change the observed
photoluminescence.\citep{Coles2014} In our work, we obtained rates
that sum across the whole polariton band in the FBZ; in practice,
experiments probe polariton photoluminescence around a narrow window
of wavevectors close to the anticrossing ($\sim10^{6}-10^{7}\text{ m}^{-1}$
in \citep{Bellessa2004}) accounting for a small fraction ($\sim$0.1
\%) of the states in the FBZ. If we take this experimental detail
into account, we notice good agreements with our theory (see Table
\ref{tbl:compare}). As an aside, we note that there are other experimental
subtleties, notably competing processes such as cavity leakage (\textasciitilde{}100-1000
fs for microcavity experiments),\citep{Lidzey2002,Coles2011PRB,Somaschi2011}
that we have not considered but may influence the observation of the
EET phenomena predicted throughout this work for the two cases of
SC.

Given the significant differences between the microcavity-\citep{Skolnick1998,Holmes2007}
and SP-based\citep{Torma2015} systems, let alone experimental uncertainty,
the accordance between our theory and the aforementioned experiments
highlights the remarkable robustness of cavity strong-coupling of
donor and acceptor excitons as a method for PARET. Moreover, we have
arrived at the third main conclusion of this paper: \textit{when donor
	and acceptors are both strongly coupled to a photonic mode, efficient
	energy exchange over hundreds of nm can occur via vibrational relaxation;
	more generally, local vibrational couplings can induce nonlocal transitions
	given sufficient delocalization of the polariton species\textemdash{}
	irrespective of spatial separation}. Seemingly ``spooky'', this
action at a (far) distance is a manifestation of donor-acceptor entanglement
resulting from strong light-matter coupling.\citep{Garcia-Vidal2017}
While this relaxation mechanism and entanglement is present in typical
molecular aggregates,\citep{KenkreBook,Sarovar2010} the novelty in
the polariton setup is the remarkable mesoscopic range of interactions
that are effectively produced.

\begin{table*}[p]
	\caption{Comparison between PARET rates for donor and acceptor cyanine dye
		J-aggregates strongly coupled to SP (theory) and microcavity (experiment)
		modes.}
	\label{tbl:compare}\centering %
	\begin{tabular}{>{\centering}m{2.5cm}>{\centering}m{2.5cm}>{\centering}m{2.5cm}>{\centering}m{2.5cm}}
		\hline 
		$\gamma$ & SP 
		
		(theory)$^{a}$ & SP
		
		(theory;
		
		experimental
		
		polariton bands)$^{b}$ & Microcavity 
		
		(experiment)$^{c}$\tabularnewline
		\hline 
		UP$\rightarrow\text{dark}_{D}$ & (10 fs$)^{-1}$ & (10 fs$)^{-1}$ & (34 fs$)^{-1}$\tabularnewline
		$\text{dark}_{D}\rightarrow$MP & (1 ps$)^{-1}$  & (1000 ps$)^{-1}$  & (603 ps$)^{-1}$\tabularnewline
		MP$\rightarrow\text{dark}_{A}$ & (10-100 fs$)^{-1}$ $^{d}$ & (10-100 fs$)^{-1}$ $^{d}$ & (8.5 fs$)^{-1}$\tabularnewline
		$\text{dark}_{A}\rightarrow$LP & (1 ps$)^{-1}$  & (1000 ps$)^{-1}$  & (228 ps$)^{-1}$\tabularnewline
		\hline 
	\end{tabular}
	
	$^{a}$These orders of magnitudes represent the ranges of rates shown
	in Fig. \ref{fgr:pda_d_fixd}b. $^{b}$Rates accounting for the fact
	that in typical polariton photoluminescence experiments, only a small
	fraction ($\sim0.1$\% of wavevectors in the FBZ) of final polariton
	states near the anticrossing is probed.\citep{Bellessa2004} $^{c}$Photoluminescence-fitted
	rate constants describing the PARET processes for a blend of J-aggregating
	NK-2707 (donors) and TDBC (acceptors) cyanine dyes both strongly coupled
	to a microcavity mode.\citep{Coles2014} $^{d}$The corresponding
	rate (see Fig. \ref{fgr:pda_d_fixd}b) spans two order of magnitudes.
\end{table*}

\section{Conclusions}

In summary, we have theoretically calculated experimentally consistent
rates of PARET for various cases of SC. We employed a polariton (plexciton)
setup consisting of a metal whose SP modes couple to donor and/or
acceptor chromophores. For strongly coupling a single type of chromophore
to SPs, we have demonstrated that energy transfer starting from delocalized
states can be enhanced due to increased spectral overlap compared
to the bare FRET case. Astonshingly, this transfer can remain fast
up to 1 $\mu\text{m}$ due to slowly decaying PRET with respect to
metal-chromophore separation when compared to the faster decaying
interchromophoric dipole-dipole coupling. Also, we have shown that
delocalizing the acceptors is a poor strategy to enhance EET starting
from the donors, but can lead to an intriguing and efficient role
reversal (carnival effect) when starting from the acceptors. These
observations shed new light on the timely debate of how to harness
coherence to enhance molecular processes.\citep{Scholes2017} Given
their generality, they can also be applied to guide the design of
polaritonic systems to control other chemical processes that have
similar donor-acceptor flavor (\emph{e.g.}, cis-trans isomerization,\citep{Galego2016,Martinez-Martinez2017}
charge transfer,\citep{Flick2017} dissociation,\citep{Flick2017}
electron transfer,\citep{Herrera2016} singlet fission\citep{Martinez-Martinez2017SF}).
Finally, our calculated rates support vibrational relaxation as the
mechanism of PARET when both donors and acceptors are strongly coupled
to a cavity mode. The results obtained in this work affirm light-matter
SC as a promising and novel means to engineer novel interactions between
molecular systems across mesoscopic lengthscales, thus opening doors
to remote controlled chemistry. 

\section*{Conflict of interest}
There are no conflicts to declare.

\section*{Acknowledgements}

M. D., R. F. R., and J. Y.-Z. acknowledge funding from the NSF CAREER
Award CHE-164732. L. A. M.-M. was supported by the UC-Mexus CONACyT
scholarship for doctoral studies. M. D.
thanks Jorge Campos-Gonz\'alez-Angulo and Rahul Deshmukh for useful
discussions. 


\balance



\providecommand*{\mcitethebibliography}{\thebibliography}
\csname @ifundefined\endcsname{endmcitethebibliography}
{\let\endmcitethebibliography\endthebibliography}{}

\clearpage

\fancypagestyle{plain}{
	\fancyhf{}
	\fancyfoot[C]{\thepage}
}
\pagestyle{plain}


\setcounter{figure}{0} \renewcommand{\thefigure}{S\arabic{figure}}
\setcounter{page}{1} \renewcommand{\thepage}{S\arabic{page}}
\setcounter{section}{0} \renewcommand{\thesection}{S\arabic{section}}
\setcounter{equation}{0} \renewcommand{\theequation}{S\arabic{equation}}
\onecolumn
\date{}

\title{Electronic Supplementary Information for ``Theory of polariton-assisted
remote energy transfer''}

\author{Matthew Du,$^{a}$ Luis A. Mart\'inez-Mart\'iinez,$^{a}$ Raphael
F. Ribeiro,$^{a}$ Zixuan Hu,$^{b}$$^{c}$ Vinod M. Menon,$^{d}$$^{e}$
and Joel Yuen-Zhou$^{\ast}$$^{a}$}
\maketitle

$^{a}$Department of Chemistry and Biochemistry, University of California
San Diego, La Jolla, California 92093, United States. \\
\\
$^{b}$Department of Chemistry, Department of Physics, and Birck
Nanotechnology Center, Purdue University, West Lafayette, IN 47907,
United States.\\
\\
$^{c}$Qatar Environment and Energy Research Institute, College of
Science and Engineering, HBKU, Doha, Qatar.\\
\\
$^{d}$Department of Physics, City College, City University of New
York, New York 10031, United States.\\
\\
$^{e}$Department of Physics, Graduate Center, City University of
New York, New York 10016, United States.\\
\\
$^{\ast}$To whom correspondence should be addressed. E-mail: \href{mailto:jyuenzhou@ucsd.edu}{jyuenzhou@ucsd.edu}

\newpage{}

\section{SP modes: additional details \label{sec:spcoupling}}

Eq. (\ref{eq:H_P sys}) and (\ref{eq:sp coupling}) in the main text
correspond to Hamiltonians representing the SP modes and their coupling
with excitons, respectively. SP modes emerge at the interface of a
metal ($z<0$ ) and dielectric medium $(z>0)$ and are assumed to
be infinitely delocalized in the $xy$-plane (in practice, they are
localized up to a coherence length, but this detail is not important
for the time being). For simplicity, the dielectric medium\textemdash which
contains both donor and acceptor slabs\textemdash is taken to have
the same real-valued positive dielectric constant $\epsilon_{d}$
at all points within. The metal is assumed to have a Drude permittivity
$\epsilon_{m}(\omega)=\epsilon_{\infty}-\frac{\omega_{P}^{2}}{\omega^{2}+i\omega\gamma}$,
where $\omega_{P}$ is the plasma frequency, $\epsilon_{\infty}$
is the high-frequency metal permittivity, and $\gamma$ is the damping
constant; we set $\gamma\to0$ in this formalism to get lossless modes
and assume that coupling of SPs to environments that induce their
relaxation processes is caused by $H_{P}^{(sys-B)}$ (see Eq. (\ref{eq:H_P sys-B})).
Here are additional definitions for terms defined in Eqs. (\ref{eq:H_P sys})
and (\ref{eq:sp coupling}) for each SP mode of in-plane momentum
$\vec{k}$ \cite{NovotnyBook2nd}:
\begin{itemize}
\item The SP mode frequency is $\omega_{\vec{k}}=|\vec{k}|c\sqrt{\frac{\epsilon_{d}+\epsilon_{m}(\omega)}{\epsilon_{d}\epsilon_{m}(\omega)}}$,
where $c$ is the speed of light in vacuum. 
\item The evanescent decay constants for the dielectric medium and metal
are given by $a_{j\vec{k}}=\sqrt{|\vec{k}|^{2}-\epsilon_{j}(\omega_{\vec{k}})\;(\omega_{\vec{k}}/c)^{2}}$,
where $j=d,\,m$.
\item The $\vec{k}$-dependent quantization length is $L_{\vec{k}}=\frac{-\epsilon_{m}}{|a_{d\vec{k}}|}+\frac{1}{2|a_{m\vec{k}}|}\left[\frac{d(\omega\epsilon_{m}(\omega))}{d\omega}\Biggr|_{\omega=\omega_{\vec{k}}}\left(\frac{\epsilon_{m}-\epsilon_{d}}{\epsilon_{m}}\right)-\epsilon_{m}-\epsilon_{d}\right]$.
It was previously reported in the literature \cite{Archambault2010}
with an additional factor of $\frac{1}{2}$; we believe our displayed
expression here is correct (see SI in \cite{Yuen-Zhou2016}).
\end{itemize}
The formalism associated with Eq. (\ref{eq:sp coupling}) has been
previously utilized in other contexts.\cite{Gonzalez-Tudela2013,Yuen-Zhou2016,Martinez-Martinez2017,Yuen-Zhou2017}

\section{Case $i$}

\subsection{Derivation of EET rate Eq. (\ref{eq:rate general case 1}) \label{subsec:Derivation:-EET-rate}}

Here we derive Eq. (\ref{eq:rate general case 1}) of the main text
by following works of Sumi\cite{Sumi1999} and Cao.\cite{Ma2015JCP1}
This formula describes the rate of transfer from eigenstate $|I\rangle$
to eigenstate $|F\rangle$ of $H_{sys}^{(i)}$ due to perturbation
$V^{(i)}$ (Section \ref{subsec:case1}), where $|I\rangle$ and $|F\rangle$
are excitonic/polaritonic states for the donor and acceptor, respectively. 

For notational simplicity, we define for this derivation $H_{D}'$
and $H_{A}'$ to be the donor and acceptor Hamiltonian, respectively,
such that $H_{C}'=H_{C}+H_{P}+H_{CP}$ for the chromophore $C$ that
is strongly coupled to a SP mode and $H_{C'}'=H_{C'}$ for the other
chromophore $C'$, which is weakly coupled to all SP modes. We also
introduce the definitions $H_{C}'^{(sys/B)}=H_{C}{}^{(sys/B)}+H_{P}^{(sys/B)}$
and $H_{C'}'^{(sys/B)}=H_{C'}^{(sys/B)}$. Then we note for clarity
that $H_{A}'^{(sys)}|I\rangle=H_{D}'^{(sys)}|F\rangle=0$. In addition,
$H_{0}^{(i)}=H_{D}'+H_{A}'$ has eigenstates $\{|\mathcal{D}\rangle\}$
($\{|\mathcal{A}\rangle\}$) that are also eigenstates of $H_{D}'$
($H_{A}'$) and whose excitonic/polaritonic-system component represents
\textit{only} donor (acceptor) but bath component represents \textit{all}
of donor, acceptor, and SPs.

Assuming excitation of donor followed by thermal equilibration occurs
before EET, the initial density matrix of the system and bath is assumed
to be $\rho_{0}=\rho_{D}^{(sys-B)}\rho_{A}^{(B)}=\sum_{\mathcal{D}}p_{0,\mathcal{D}}|\mathcal{D}\rangle\langle\mathcal{D}|,$
where $p_{0,\mathcal{D}}$ is the equilibrium occupation probability
of system-bath state $|\mathcal{D}\rangle$. The constituent density
matrices $\rho_{D}^{(sys-B)}=\frac{e^{-H_{D}'/k_{B}T}}{\text{tr}e^{-H_{D}'/k_{B}T}}$
and $\rho_{A}^{(B)}=\frac{e^{-H_{A}'^{(B)}/k_{B}T}}{\text{tr}e^{-H_{A}'^{(B)}/k_{B}T}}$
respectively represent the pre-EET equilibrium distribution of donor
excited vibronic and acceptor ground vibrational states, in corresponding
tensor products with photon states. Then the rate of transfer from
$|I\rangle$ to $|F\rangle$ given by Fermi's golden rule is
\begin{equation}
\gamma_{F\leftarrow I}=\frac{2\pi}{\hbar}\sum_{\mathcal{D},\mathcal{A}}p_{0,\mathcal{D}}|\langle\mathcal{P}_{F}\mathcal{A}|V^{(i)}|\mathcal{P}_{I}\mathcal{D}\rangle|^{2}\delta(\hbar\omega_{\mathcal{A},\mathcal{D}}),\label{eq:fgr initial}
\end{equation}
The state $|\mathcal{P}_{I}\mathcal{D}\rangle$ ($|\mathcal{P}_{F}\mathcal{A}\rangle$)
results from projection $\mathcal{P}_{I}=|I\rangle\langle I|$ ($\mathcal{P}_{F}=|F\rangle\langle F|$)
of the system component of $|\mathcal{D}\rangle$ ($|\mathcal{A}\rangle$)
onto $|I\rangle$ ($|F\rangle$). To express this equation in the
(purely electronic/polaritonic) eigenbasis $\{|D\rangle\}\cup\{|A\rangle\}$
of $H_{sys}^{(i)}$, we first use the fact that the bath modes on
each type of chromophore are independent (see Eqs. (\ref{eq:xb})
and (\ref{eq:xeb})) and write the time-domain expression:
\begin{align}
\gamma_{F\leftarrow I} & =\frac{1}{\hbar^{2}}\int_{-\infty}^{\infty}dt\sum_{\mathcal{D},\mathcal{A}}p_{0,\mathcal{D}}\langle\mathcal{P}_{I}\mathcal{D}|V^{(i)}|\mathcal{P}_{F}\mathcal{A}\rangle\langle\mathcal{P}_{F}\mathcal{A}|e^{-iH_{0}^{(i)}t/\hbar}V^{(i)}e^{iH_{0}^{(i)}t/\hbar}|\mathcal{P}_{I}\mathcal{D}\rangle\nonumber \\
 & =\frac{1}{\hbar^{2}}\int_{-\infty}^{\infty}dt\sum_{\mathcal{D}}p_{0,\mathcal{D}}\langle\mathcal{P}_{I}\mathcal{D}|V^{(i)}\mathcal{P}_{F}e^{-i(H_{A}'+H_{D}'^{(B)})t/\hbar}V^{(i)}e^{i(H_{D}'+H_{A}'^{(B)})t/\hbar}|\mathcal{P}_{I}\mathcal{D}\rangle,
\end{align}
where in the second line, we have used $\sum_{\mathcal{A}}|\mathcal{P}_{F}\mathcal{A}\rangle\langle\mathcal{P}_{F}\mathcal{A}|e^{-iH_{0}^{(i)}t/\hbar}=\mathcal{P}_{F}e^{-i(H_{A}'+H_{D}'^{(B)})t/\hbar}$
and $e^{iH_{0}^{(i)}t/\hbar}|\mathcal{P}_{I}\mathcal{D}\rangle=e^{i(H_{D}'+H_{A}'^{(B)})t/\hbar}|\mathcal{P}_{I}\mathcal{D}\rangle$.
Using the independence of the $D,A$ vibrational bath modes, we obtain
\begin{align}
\gamma_{F\leftarrow I} & =\frac{1}{\hbar^{2}}\int_{-\infty}^{\infty}dt\sum_{D'}\sum_{A'}V_{IF}^{(i)}V_{D'A'}^{(i)*}\text{tr}_{bD}\{e^{-iH_{D}'^{(B)}t/\hbar}\langle D'|e^{iH_{D}'t/\hbar}\mathcal{P}_{I}\rho_{D}^{(sys-B)}|D\rangle\}\text{tr}_{bA}\{e^{iH_{A}'^{(B)}t/\hbar}\rho_{A}^{(B)}\langle F|e^{-iH_{A}'t/\hbar}|A'\rangle\}
\end{align}
for $V_{D'A'}^{(i)}=\langle D'|V^{(i)}|A'\rangle$; here, $\text{tr}_{bC}$
denotes a trace over the bath degrees of freedom associated with $C$.
In the limit of weak exciton-bath coupling, $[e^{iH_{D}'t/\hbar},\mathcal{P}_{I}]\approx0$,
and $\langle D'|e^{iH_{D}t/\hbar}\rho_{D}^{(sys-B)}|I\rangle\approx0$
($\langle F|e^{-iH_{A}t/\hbar}|A'\rangle\approx0$) for $D'\neq I$
($A'\neq F$). Then
\begin{equation}
\gamma_{F\leftarrow I}\approx\frac{1}{\hbar^{2}}|V_{FI}^{(i)}|^{2}\int_{-\infty}^{\infty}dt\,\text{tr}_{bD}\{e^{-iH_{D}'^{(B)}t/\hbar}\langle I|e^{iH_{D}'t/\hbar}\rho_{D}^{(sys-B)}|I\rangle\}\text{tr}_{bA}\{e^{iH_{A}'^{(B)}t/\hbar}\rho_{A}^{(B)}\langle F|e^{-iH_{A}'t/\hbar}|F\rangle\}.\label{eq:gamma_FI}
\end{equation}
To proceed further, let us define the operators
\begin{align}
\mathcal{I}_{F}(\omega) & =\frac{1}{2\pi}\int_{-\infty}^{\infty}dt\,e^{i\omega t}\text{tr}_{bA}\{e^{iH_{A}'^{(B)}t/\hbar}\rho_{A}^{(B)}\langle F|e^{-iH_{A}'t/\hbar}|F\rangle\},\label{eq:abs formal}\\
\mathcal{E}_{I}(\omega) & =\frac{1}{2\pi}\int_{-\infty}^{\infty}dt\,e^{-i\omega t}\text{tr}_{bD}\{e^{-iH_{D}'^{(B)}t/\hbar}\langle I|e^{iH_{D}'t/\hbar}\rho_{D}^{(sys-B)}|I\rangle\},\label{eq:em formal}
\end{align}
representing the spectra for absorption of acceptor state $F$ and
emission of donor state $I$, respectively. Also, denote the corresponding
spectral overlap as 
\begin{equation}
J_{F,I}=\frac{1}{\hbar}\int_{-\infty}^{\infty}d\omega\,\mathcal{E}_{I}(\omega)\mathcal{I}_{F}(\omega).
\end{equation}
Eq. (\ref{eq:gamma_FI}) then yields,
\begin{align}
\gamma_{F\leftarrow I} & \approx\frac{2\pi}{\hbar}|V_{FI}^{(i)}|^{2}J_{F,I},\label{eq:preJ rate}
\end{align}
which is Eq. (\ref{eq:rate general case 1}) of the main text; in
other words, the spectral overlap $J_{F,I}$ takes on the role of
a density of final states for a Fermi golden rule rate. We note that
for our simulations where the donors and acceptors are reversed (see
Sections \ref{sec:application} and \ref{subsec:rates reversed}),
Eq. (\ref{eq:preJ rate}) still applies except $D,I$ ($A,F$) now
refer to acceptors (donors). 

\subsection{Estimation of spectral overlaps \label{subsec:lorentzian }}

As an illustration of the application of our theory, in this subsection,
we develop a simplified Lorentzian overlap model to compute $J_{F,I}$
at temperature $T=0$. To write Eqs. (\ref{eq:abs formal}) and (\ref{eq:em formal})
as Lorentzian lineshapes, we follow a Feshbach projection operator
approach.\cite{Feshbach1958paper1,Feshbach1962paper2,MukamelBook,ShapiroBook}
While we proceed with the specific case of only donors strongly coupled
to a SP (\textit{i.e.}, $H_{A}'=H_{A}$ and $H_{D}'=H_{D}+H_{P}+H_{DP}$),
the derivation can be readily extended to the case of acceptors strongly
coupled to a SP. 

Define the projectors 
\begin{align}
\mathcal{P}_{A} & =\sum_{l,m}|A_{lm}\rangle\langle A_{lm}|\otimes|0_{B}\rangle\langle0_{B}|,\\
\mathcal{P}_{D} & =\sum_{i,j}|D_{ij}\rangle\langle D_{ij}|\otimes|0_{B}\rangle\langle0_{B}|+\sum_{\vec{k}\in\text{FBZ}}|\vec{k}\rangle\langle\vec{k}|\otimes|0_{B}\rangle\langle0_{B}|
\end{align}
($|0_{B}\rangle$ is the vacuum state for chromophores and SP bath
modes, and $|\vec{k}\rangle=a_{\vec{k}}^{\dagger}|0\rangle$), which
map vibronic (possibly including photonic component) states onto electronic/photonic\textemdash \emph{i.e.},
``bathless''\textemdash states. Let the identity on the entire set
of degrees of freedom be $1_{sys-B}$ and define the projectors $\mathcal{Q}_{A}=1_{sys-B}-\mathcal{P}_{A}$
and $\mathcal{Q}_{D}=1_{sys-B}-\mathcal{P}_{D}$. Then the time-dependent
Schrödinger equation $\frac{d}{dt}|\psi(t)\rangle=-\frac{i}{\hbar}H_{C}'|\psi(t)\rangle$
for $C=D,A$ can be rewritten as the coupled equations

\begin{subequations}\label{eq:coupled}
\begin{align}
-\frac{i}{\hbar}\mathcal{P}_{C}H_{C}^{'}\mathcal{P}_{C}|\mathcal{P}_{C}\psi(t)\rangle-\frac{i}{\hbar}\mathcal{P}_{C}H_{C}^{'}\mathcal{Q}_{C} & |\mathcal{Q}_{C}\psi(t)\rangle=\frac{d}{dt}|\mathcal{P}_{C}\psi(t)\rangle,\label{eq:peq}\\
-\frac{i}{\hbar}\mathcal{Q}_{C}H_{C}^{'}\mathcal{P}_{C}|\mathcal{P}_{C}\psi(t)\rangle-\frac{i}{\hbar}\mathcal{Q}_{C}H_{C}^{'}Q & _{C}|\mathcal{Q}_{C}\psi(t)\rangle=\frac{d}{dt}|\mathcal{Q}_{C}\psi(t)\rangle.\label{eq:qeq}
\end{align}

\end{subequations} \noindent After the initial condition $|\mathcal{Q}_{C}\psi(0)\rangle=0$,
multiplying both sides of Eq. (\ref{eq:coupled}) by $\Theta(t)$,
and plugging the formal solution of Eq. (\ref{eq:qeq}) into Eq. (\ref{eq:peq}),
we obtain the Green functions $G_{C}(t)=\Theta(t)e^{-iH_{C}'t/\hbar}$
which can be Fourier transformed as,\cite{MukamelBook}

\begin{subequations}\label{eq:Gs1}
\begin{align}
G_{A}(\hbar\omega) & =-i\int_{-\infty}^{\infty}dt\,e^{i\omega t}G_{A}(t)=\text{lim}_{\epsilon\to0^{+}}\frac{1}{\hbar\omega-H_{A}^{(sys)}\otimes|0_{B}\rangle\langle0_{B}|-R_{A}(\hbar\omega)+i\epsilon},\label{eq:g A}\\
G_{D}(\hbar\omega) & =-i\int_{-\infty}^{\infty}dt\,e^{i\omega t}G_{D}(t)=\text{lim}_{\epsilon\to0^{+}}\frac{1}{\hbar\omega-(H_{D}^{(sys)}+H_{P}^{(sys)}+H_{DP})\otimes|0_{B}\rangle\langle0_{B}|-R_{DP}(\hbar\omega)+i\epsilon},\label{eq:g D}
\end{align}

\end{subequations} \noindent where

\begin{subequations}\label{eq:R}
\begin{align}
R_{A}(\hbar\omega) & =\mathcal{P}_{A}H_{A}^{(sys-B)}\mathcal{Q}_{A}\frac{1}{\hbar\omega-\mathcal{Q}_{A}H_{A}'\mathcal{Q}_{A}+i\epsilon}\mathcal{Q}_{A}H_{A}^{(sys-B)}\mathcal{P}_{A},\label{eq:r self energy a}\\
R_{D}(\hbar\omega) & =\mathcal{P}_{D}[H_{D}^{(sys-B)}+H_{P}^{(sys-B)}]\mathcal{Q}_{D}\frac{1}{\hbar\omega-\mathcal{Q}_{D}H_{D}'\mathcal{Q}_{D}+i\epsilon}\mathcal{Q}_{D}[H_{D}^{(sys-B)}+H_{P}^{(sys-B)}]\mathcal{P}_{D},\label{eq:r self energy d}
\end{align}

\end{subequations} \noindent are the so-called self-energy terms.
We next make the partition $R_{C}(\hbar\omega)=\text{Re}R_{C}(\hbar\omega)-\frac{i}{2}\Gamma_{C}(\hbar\omega)$
with $\Gamma_{C}(\hbar\omega)=-2\text{Im}R_{C}(\hbar\omega)$ and
assume that the Lamb shifts $\text{Re}R_{C}(\hbar\omega)$ contribute
insignificantly when compared to the exciton/SP energies and can thus
be neglected. In addition, we take the wide-band approximation that
all (diagonal) matrix elements of $\Gamma_{C}(\hbar\omega)$ are constant
as a function of $\omega$: $\langle A_{lm},0_{B}|\Gamma_{A}(\hbar\omega)|A_{lm},0_{B}\rangle\approx\Gamma_{A}$,
$\langle D_{ij},0_{B}|\Gamma_{D}(\hbar\omega)|D_{ij},0_{B}\rangle\approx\Gamma_{D}$,
and $\langle1_{\vec{k}},0_{B}|\Gamma_{D}(\hbar\omega)|1_{\vec{k}},0_{B}\rangle\approx\Gamma_{P,\vec{k}}$.
Then explicitly writing out the respective Hamiltonians in Eq. (\ref{eq:Gs1}),
we arrive at

\begin{subequations}\label{eq:Gs}
\begin{align}
G_{A}(\hbar\omega) & =\text{lim}_{\epsilon\to0^{+}}\frac{1}{\hbar\omega-(\hbar\omega_{A}-\frac{i}{2}\Gamma_{A})\sum_{l,m}|A_{lm}\rangle\langle A_{lm}|\otimes|0_{B}\rangle\langle0_{B}|+i\epsilon}\\
G_{D}(\hbar\omega) & =\text{lim}_{\epsilon\to0^{+}}\frac{1}{\hbar\omega-[(\hbar\omega_{D}-\frac{i}{2}\Gamma_{D})\sum_{i,j}|D_{ij}\rangle\langle D_{ij}|+\sum_{\vec{k}\in\text{FBZ}}(\hbar\omega_{\vec{k}}-\frac{i}{2}\Gamma_{P,\vec{k}})a_{\vec{k}}^{\dagger}a_{\vec{k}}+H_{DP}]\otimes|0_{B}\rangle\langle0_{B}|+i\epsilon}\label{eq:Gd prediag}\\
 & =\text{lim}_{\epsilon\to0^{+}}\frac{1}{\hbar\omega-[(\hbar\omega_{D}-\frac{i}{2}\Gamma_{D})\sum_{\vec{k}\in\text{FBZ}}\sum_{d}|d_{\vec{k}}\rangle\langle d_{\vec{k}}|+\sum_{\vec{k}\in\text{FBZ}}\sum_{\alpha}(\hbar\omega_{\alpha_{D,\vec{k}}}-\frac{i}{2}\Gamma_{\alpha_{D,\vec{k}}})|\alpha_{D,\vec{k}}\rangle\langle\alpha_{D,\vec{k}}|]\otimes|0_{B}\rangle\langle0_{B}|+i\epsilon}.\label{eq:Gd diag}
\end{align}

\end{subequations} \noindent It is intuitively clear that the terms
in brackets in the denominators of Eqs. (\ref{eq:Gd prediag}) and
(\ref{eq:Gd diag}) correspond to effective Hamiltonians of the donor-SP
coupled system. In fact, $\hbar\omega_{\alpha_{D,\vec{k}}}$ and $\Gamma_{\alpha_{D,\vec{k}}}$
are the resulting (real-valued) energy and linewidth of $|\alpha_{D,\vec{k}}\rangle$
(defined just after Eq. (\ref{eq:polariton1}) in the main text) for
$\alpha=\text{UP},\text{LP}$. Comparing those two equations shows
that not only can the polariton energies and linewidths be obtained
from the diagonalization of a non-Hermitian Hamiltonian for the donors
and SP modes, but all dark states have the same energy and linewidth
as the bare chromophores. 

More precisely, applying the assumption of weak system-bath coupling
that was used to derive Eq. (\ref{eq:rate general case 1}) and setting
$\rho_{C}^{(B)}=|0_{B}\rangle\langle0_{B}|$ ($T=0$), we can readily
relate the expressions for absorption and emission spectra (Eqs. (\ref{eq:abs formal})
and (\ref{eq:em formal}), respectively) to matrix elements of the
Green's functions (see Eq. (\ref{eq:Gs1})),

\begin{subequations}\label{eq:absem approx}

\begin{align}
\mathcal{I}_{F}(\omega) & \approx\frac{1}{2\pi}\text{tr}_{bA}\{e^{iH_{A}^{(B)}t/\hbar}\rho_{A}^{(B)}\langle F|-2\text{Im}G_{A}(\hbar\omega)|F\rangle\}\nonumber \\
 & =\frac{1}{\pi}\frac{\frac{1}{2}\Gamma_{F}}{(\hbar\omega-\hbar\omega_{I})^{2}+(\frac{1}{2}\Gamma_{F})^{2}},\\
\mathcal{E}_{I}(\omega) & \approx\frac{1}{2\pi}\text{tr}_{bD}\{e^{-iH_{D}^{(B)}t/\hbar}\rho_{D}^{(B)}\langle I|-2\text{Im}G_{D}(\hbar\omega)|I\rangle\}\nonumber \\
 & =\frac{1}{\pi}\frac{\frac{1}{2}\Gamma_{I}}{(\hbar\omega-\hbar\omega_{I})^{2}+(\frac{1}{2}\Gamma_{I})^{2}},
\end{align}

\end{subequations} \noindent  with $F=A$ ($I=\alpha_{D,\vec{k}},D$)
representing any state of type $|A_{lm}\rangle$ ($|\alpha_{\vec{k},D}\rangle,|d_{\vec{k}}\rangle$),
and we thus obtain the Lorentzian spectral overlap $J_{F,I}=\frac{1}{\pi}\frac{\frac{\Gamma_{I}+\Gamma_{F}}{2}}{(\frac{\Gamma_{I}+\Gamma_{F}}{2})^{2}+(\hbar\omega_{FI})^{2}}$.

Although we made several approximations to achieve this form, Lorentzian
behavior is usually characteristic of lineshapes near their peaks,\cite{Cohen2003}
leading to overlaps and thus EET rates that are especially relevant
when initial and final states are near resonance. In passing, we mention
that a different physical mechanism to obtain Lorentzian lineshapes
occurs when electronic states are coupled to overdamped Brownian oscillators
in the limits of high temperature and fast nuclear dynamics\cite{MukamelBook}. 

\subsection{Lack of supertransfer enhancement \label{subsec:no superradiant enhancement}}

Here, we demonstrate that even though the donor polariton state is
coherently delocalized, the superradiance enhancement of EET to bare
acceptors that one could expect\cite{Lloyd2010,Kassal2013} is negligible
when taking into account the distance dependence of the involved dipolar
interactions. The essence of supertransfer is that a constructive
interference of individual donor dipoles in an aggregate can lead
to FRET rates that scale as $N$ times a bare FRET rate. However,
for this to happen, it is important to have a geometric arrangement
where all donors are equidistantly spaced with respect to all acceptors;
this is not the case in our problem. 

To show this point explicitly, we first evaluate the FRET rate associated
with the delocalized donor $|D_{\vec{k}=0}\rangle$ transferring energy
to bare acceptors,

\begin{align}
\gamma_{A\leftarrow D_{\vec{k}=0}}^{FRET} & \equiv\frac{2\pi}{\hbar}\sum_{l,m}|\langle A_{lm}|H_{DA}|D_{\vec{k}=\vec{0}}\rangle|^{2}J_{A,D_{\vec{k}=\vec{0}}}\nonumber \\
 & \approx\frac{2\pi}{\hbar}\mu_{D}^{2}\mu_{A}^{2}\frac{1}{N_{D}}\sum_{l,m}\left|\sum_{i,j}\frac{\kappa_{ijlm}}{r_{ijlm}^{3}}\right|^{2}J_{A,D_{\vec{k}=\vec{0}}},\label{eq:prerad}
\end{align}
where we have approximated $|D_{\vec{k}=0}\rangle$ as a totally symmetric
state across all chromophores, $|c_{D_{i'j'}D_{\vec{k}=0}}|^{2}=\frac{\kappa_{\vec{k}D_{i'j'}}^{2}}{\sum_{i,j}\kappa_{\vec{k}D_{ij}}^{2}}\approx\frac{1}{N_{D}}$
(see definition of $|C_{\vec{k}}\rangle$ for $C=D,A$ in main text
right after Eq. (\ref{eq:polariton1})). 

Compare Eq. (\ref{eq:prerad}) to the corresponding bare FRET rate
\begin{equation}
\gamma_{\text{bare FRET}}''=\frac{2\pi}{\hbar}\mu_{D}^{2}\mu_{A}^{2}\frac{1}{N_{D}}\sum_{l,m}\sum_{i,j}\frac{\kappa_{ijlm}^{2}}{r_{ijlm}^{6}}J_{A,D}.\label{eq:bare FRET 3}
\end{equation}
Since the separation between donor and acceptor slabs is small compared
to their longitudinal ($xy$) dimensions, it is incorrect to approximate
the distance $r_{ijlm}$ as constant for all $i,j,l,m$. To see this
more precisely, 

\begin{align}
\left|\sum_{i,j}\frac{\kappa_{ijlm}}{r_{ijlm}^{3}}\right|^{2} & \leq4\left(\sum_{i,j}\frac{1}{r_{ijlm}^{3}}\right)^{2}\nonumber \\
 & =4\left(\rho_{D}\int_{s}^{s+\mathcal{W}}dz\int d\vec{R}\frac{1}{[|\vec{R}-\vec{R}_{l,A}|^{2}+(z-z_{m,A})^{2}]^{\frac{3}{2}}}\right)^{2}\nonumber \\
 & =4\left(\rho_{D}\int_{s}^{s+\mathcal{W}}dz\int_{0}^{2\pi}d\phi\int_{0}^{\infty}dr\frac{r}{[r^{2}+(z-z_{m,A})^{2}]^{\frac{3}{2}}}\right)^{2}\nonumber \\
 & \ll\lim_{N_{D}\rightarrow\infty}N_{D}\sum_{i,j}\frac{\kappa_{ijlm}^{2}}{r_{ijlm}^{6}}.
\end{align}
The finiteness of both the concentration $\rho_{D}$ and the integral
over $(r,\phi,z)$ allow going from the third to the fourth line as
long as $N_{D}$ is sufficiently large. More precisely, the integral
in the third line with respect to $r$ scales as $r^{-1}$ and converges
for $z_{m,A}>z_{j,D}$ for all $j$; the sum $\sum_{i,j}\frac{\kappa_{ijlm}^{2}}{r_{ijlm}^{6}}$
in the fourth line can be similarly converted into an integral which
converges as $r^{-4}$. We have also used the fact that the squared
FRET orientation factor $\kappa_{ijlm}^{2}$ ranges from 0 to 4.\cite{MedintzBook}
For large enough $N_{D}$, the difference in spectral overlaps $J_{A,D}$
and $J_{A,D_{\vec{k}=\vec{0}}}$ is immaterial. Therefore, if $\gamma_{\text{bare FRET}}''\neq0$,
$\gamma_{A\leftarrow D_{\vec{k}=0}}^{FRET}\ll N_{D}\gamma_{\text{bare FRET}}''$
and we conclude that the decay of dipolar interactions with respect
to distance precludes a supertransfer enhancement\cite{Lloyd2010,Kassal2013}
in our problem. 

By noticing that $\left|\sum_{i,j}\frac{\kappa_{ijlm}}{r_{ijlm}^{3}}\right|^{2}>\sum_{i,j}\frac{\kappa_{ijlm}^{2}}{r_{ijlm}^{6}}$
(at least when all $\kappa_{ijlm}\geq0$), and for $J_{A,D}\approx J_{A,D_{\vec{k}=\vec{0}}}$,
we still expect a coherence enhancement of EET: $\gamma_{A\leftarrow D_{\vec{k}=0}}^{FRET}>\gamma_{\text{bare FRET}}''$.
However, this enhancement is quite modest compared to all other effects
that we consider in our problem (\emph{e.g.}, PRET contributions). 

\subsection{Derivation of rate Eq. (\ref{eq:first rate k-1}) from donor polaritons
to acceptors\label{subsec:derivation_donor polaritons}}

Here, we derive expressions for EET rates between a multi-layer slab
of $N_{D}\gg N_{xy,D}$ donor molecules strongly coupled to a SP and
a monolayer of acceptor molecules at $z=z_{0,A}>z_{j,D}$ for all
$j$. The polariton states $\alpha=\text{UP},\text{LP}$ are of the
form $|\alpha_{D,\vec{k}}\rangle=c_{D_{\vec{k}}\alpha_{D,\vec{k}}}|D_{\vec{k}}\rangle+c_{\vec{k}\alpha_{D,\vec{k}}}|\vec{k}\rangle$,
where $|\vec{k}\rangle=a_{\vec{k}}^{\dagger}|0\rangle$ and 
\begin{equation}
|D_{\vec{k}}\rangle=\frac{1}{\sqrt{\sum_{i,j}\kappa_{\vec{k}D_{ij}}^{2}e^{-2a_{d\vec{k}}z_{j,D}}}}\sum_{i,j}\kappa_{\vec{k}D_{ij}}e^{-a_{d\vec{k}}z_{j,D}}e^{i\vec{k}\cdot\vec{R}_{i,D}}|D_{ij}\rangle.\label{eq:donor crude}
\end{equation}
Plugging Eq. (\ref{eq:donor crude}) into Eq. (\ref{eq:first rate k}),
the rate of transfer from donor polariton state $|\alpha_{D,\vec{k}}\rangle$
($\alpha=\text{UP},\text{LP}$) to acceptors is
\begin{align}
\gamma_{A\leftarrow\alpha_{D,\vec{k}}} & =\frac{2\pi}{\hbar}\sum_{l}\left|\langle A_{l0}|H_{DA}+H_{AP}|\alpha_{D}\rangle\right|^{2}J_{A\alpha_{D,\vec{k}}}\nonumber \\
 & =\frac{2\pi}{\hbar}\sum_{l}\left|\langle A_{l0}|H_{DA}+H_{AP}|\left(c_{D_{\vec{k}}\alpha_{D,\vec{k}}}\sum_{i,j}\frac{\kappa_{\vec{k}D_{ij}}e^{-a_{d\vec{k}}z_{j,D}}e^{i\vec{k}\cdot\vec{R}_{i,D}}}{\sqrt{\sum_{i,j}\kappa_{\vec{k}D_{ij}}^{2}e^{-2a_{d\vec{k}}z_{j,D}}}}|D_{ij}\rangle+c_{\vec{k}\alpha_{D,\vec{k}}}|\vec{k}\rangle\right)\right|^{2}J_{A,\alpha_{D,\vec{k}}}.\label{eq:show plex to a}
\end{align}
For simplicity, we next assume the chromophore TDMs are isotropically
distributed and orientationally uncorrelated, 

\begin{subequations}\label{eq:almost first rate k}
\begin{align}
\gamma_{A\leftarrow\alpha_{D,\vec{k}}} & =|c_{D_{\vec{k}}\alpha_{D,\vec{k}}}|^{2}\frac{2\pi}{\hbar}N_{A}\sum_{i,j}\Bigg(\frac{e^{-2a_{d\vec{k}}z_{j}}}{N_{xy,D}\sum_{j}e^{-2a_{d\vec{k}}z_{j,D}}}\Bigg)\frac{\mu_{D}^{2}\mu_{A}^{2}\langle\kappa_{FRET}^{2}\rangle}{r_{ij}^{6}}J_{A\alpha_{D,\vec{k}}}\nonumber \\
 & \qquad+|c_{\vec{k}\alpha_{D,\vec{k}}}|^{2}\frac{2\pi}{\hbar}\rho_{A}^{(2D)}\mu_{A}^{2}\langle\kappa_{LM,\vec{k}}^{2}\rangle\frac{\hbar\omega_{\vec{k}}}{2\epsilon_{0}L_{\vec{k}}}e^{-2a_{d\vec{k}}z_{0,A}}J_{A\alpha_{D,\vec{k}}},\label{eq:almost first rate k-2}
\end{align}

\end{subequations} \noindent where $r_{ij}$ is distance between
acceptor at $(0,0,z_{0,A})$ and donor $ij$, $\rho_{A}^{(2D)}=\frac{N_{xy,A}}{S}$
is the concentration of acceptors per unit area, the isotropically
averaged orientation factors for FRET and light-matter interaction
are $\langle\kappa_{FRET}^{2}\rangle=\langle\kappa_{ijl0}^{2}\rangle=\frac{2}{3}$\cite{MedintzBook},
and $\langle\kappa_{LM,\vec{k}}^{2}\rangle=\langle\kappa_{\vec{k}D_{ij}}^{2}\rangle=\frac{2}{3}+\frac{1}{3}\frac{|\vec{k}|^{2}}{a_{d\vec{k}}^{2}}$\cite{Gonzalez-Tudela2013}.
Eq. (\ref{eq:almost first rate k-2}) shows that the isotropic distribution
of dipoles and the lack of correlations amongst their orientations
yields an incoherently averaged rate over the populations of exciton
(first term) and SP (second term). Furthermore, Eq. (\ref{eq:last rate 1-1})
of the main text is a less explicit form of Eq. (\ref{eq:almost first rate k-2})
that follows from using the approximation $\frac{e^{-2a_{d\vec{k}}z_{j,D}}}{N_{xy,D}\sum_{j}e^{-2a_{d\vec{k}}z_{j,D}}}\approx|c_{D_{ij}D_{\vec{k}}}|^{2}=\frac{\kappa_{\vec{k}D_{ij}}^{2}e^{-2a_{d\vec{k}}z_{j,D}}}{\sum_{i,j}\kappa_{\vec{k}D_{ij}}^{2}e^{-2a_{d\vec{k}}z_{j,D}}}$
(\textit{i.e.}, ignoring the orientational dependence of the exciton
populations). 

In obtaining Eq. (\ref{eq:almost first rate k-2}), we have utilized
the following approximations which are valid for large $N_{D}$:

\begin{subequations}\label{eq:approx separate}
\begin{align}
\left\langle \frac{\kappa_{\vec{k}D_{ij}}\kappa_{ijl0}\kappa_{\vec{k}D_{i'j'}}\kappa_{i'j'l0}}{\sum_{i,j}\kappa_{\vec{k}D_{ij}}^{2}e^{-2a_{d\vec{k}}z_{j,D}}}\right\rangle  & \approx\frac{\langle\kappa_{\vec{k}D_{ij}}\kappa_{ijl0}\kappa_{\vec{k}D_{i'j'}}\kappa_{i'j'l0}\rangle}{\sum_{i,j}\langle\kappa_{\vec{k}D_{ij}}^{2}\rangle e^{-2a_{d\vec{k}}z_{j,D}}}=\frac{\langle\kappa_{\vec{k}D_{ij}}\kappa_{ijl0}\kappa_{\vec{k}D_{i'j'}}\kappa_{i'j'l0}\rangle}{\langle\kappa_{LM}^{2}\rangle N_{xy,D}\sum_{j}e^{-2a_{d\vec{k}}z_{j,D}}},\label{eq:1st approx orientation}\\
\left\langle \frac{\kappa_{\vec{k}D_{ij}}\kappa_{ijlm}\kappa_{\vec{k}A_{l0}}}{\sqrt{\sum_{i,j}\kappa_{\vec{k}D_{ij}}^{2}e^{-2a_{d\vec{k}}z_{j,D}}}}\right\rangle  & \approx\frac{\langle\kappa_{\vec{k}D_{ij}}\kappa_{ijlm}\kappa_{\vec{k}A_{l0}}\rangle}{\langle\sqrt{\sum_{i,j}\kappa_{\vec{k}D_{ij}}^{2}e^{-2a_{d\vec{k}}z_{j,D}}}\rangle}.\label{eq:1st approx orientation 2}
\end{align}

\end{subequations}\noindent In addition, we have applied a mean-field
approach to the orientational factors,

\begin{subequations}\label{eq:decouple}
\begin{align}
\langle\kappa_{\vec{k}D_{ij}}\kappa_{ijl0}\kappa_{\vec{k}D_{'i'j}}\kappa_{i'j'l0}\rangle & \approx\langle\kappa_{\vec{k}D_{ij}}\kappa_{\vec{k}D_{'i'j}}\rangle\langle\kappa_{ijl0}\kappa_{i'j'l0}\rangle=\langle\kappa_{LM,\vec{k}}^{2}\rangle\langle\kappa_{FRET}^{2}\rangle\delta_{(i,j),(i'j')},\label{eq:decouple1}\\
\langle\kappa_{\vec{k}D_{ij}}\kappa_{ijl0}\kappa_{\vec{k}A_{l0}}\rangle & \approx\langle\kappa_{\vec{k}D_{ij}}\kappa_{\vec{k}A_{l0}}\rangle\langle\kappa_{ijl0}\rangle=0.\label{eq:decouple2}
\end{align}

\end{subequations}\noindent In the continuum limit, Eq. (\ref{eq:almost first rate k-2})
reads,
\begin{equation}
\gamma_{A\leftarrow\alpha_{D,\vec{k}}}=\frac{2\pi}{\hbar}\rho_{A}^{(2D)}\left[|c_{D_{\vec{k}}a_{D,\vec{k}}}|^{2}\mu_{D}^{2}\mu_{A}^{2}\langle\kappa_{FRET}^{2}\rangle\frac{\int_{s}^{s+\mathcal{W}}dze^{-2a_{d\vec{k}}z}\frac{\pi}{2(z_{0,A}-z)^{4}}}{\frac{e^{-2a_{d\vec{k}}(s+\mathcal{W})}-e^{-2a_{d\vec{k}}s}}{-2a_{d\vec{k}}}}+|c_{\vec{k}\alpha_{D,\vec{k}}}|^{2}\mu_{A}^{2}\langle\kappa_{LM,\vec{k}}^{2}\rangle\frac{\hbar\omega_{\vec{k}}}{2\epsilon_{0}L_{\vec{k}}}e^{-2a_{d\vec{k}}z_{0,A}}\right]J_{A\alpha_{D,\vec{k}}},\label{eq:plextoa}
\end{equation}
where the base of the donor slab is located at $z=s$ and its thickness
is $\mathcal{W}$ (while the integral over $z$ is analytical, it
is complicated and does not shed much insight into the problem).

\subsection{Derivation of rate Eq. (\ref{eq:darkD to A-1}) from donor dark states
to acceptors\label{subsec:derivation_dark donors}}

Here, we show that the rate (Eq. (\ref{eq:darkD to A})) of EET from
donor dark states to a spatially separated monolayer of bare acceptors
at $z=z_{0,A}$ converges to the bare FRET rate $\gamma_{\text{bare FRET}}$
in Eq. (\ref{eq:darkD to A-1}) assuming that $N_{D}\gg N_{xy,D}$
and all donor and acceptor TDMs are isotropically oriented and uncorrelated.
The steps applied in this subsection can be extended in a straightforward
manner to establish this result for multiple layers in the acceptor
slab. This result is intuitively expected given that the dark states
are purely excitonic and centered at the original transition frequency
$\omega_{D}$; furthermore, the density of dark states is close to
the original density of bare donor states. Our derivation here relies
on Lorentzian lineshapes for the dark donor and acceptor states (see
Section \ref{subsec:lorentzian }); however, we anticipate the conclusions
to hold for more general lineshapes under certain limits. 

Section \ref{subsec:lorentzian } specifically reveals that weak system-bath
coupling and $T=0$ allow the lineshape of each dark state $|d_{D,\vec{k}}\rangle$
to be expressed as a Lorentzian with peak energy and linewidth identical
to that of bare donors, leading to $J_{A,d_{D,\vec{k}}}=J_{A,D}$.
Connecting this finding to the relevant rate expression, we can rewrite
Eq. (\ref{eq:darkD to A}) of the main text for an acceptor monolayer
as

\begin{align}
\gamma_{A\leftarrow dark_{D}} & =\frac{2\pi}{\hbar}\frac{1}{N_{D}-N_{xy,D}}\sum_{l}\left[\sum_{i,j}|\langle A_{l0}|H_{DA}|D_{ij}\rangle|^{2}-\sum_{\vec{k}\in\text{FBZ}}|\langle A_{l0}|H_{DA}|D_{\vec{k}}\rangle|^{2}\right]J_{A,D}.
\end{align}
We have used the relation $\sum_{\vec{k}\in\text{FBZ}}\sum_{d}|d_{\vec{k}}\rangle\langle d_{\vec{k}}|=1_{D}^{(sys)}-\sum_{\vec{k}\in\text{FBZ}}|D_{\vec{k}}\rangle\langle D_{\vec{k}}|$
for donor (electronic) identity $1_{D}^{(sys)}$. Assuming $N_{z,D}\gg1$
(equivalent to $N_{D}\gg N_{xy,D}$ ), we have 
\begin{align}
\gamma_{A\leftarrow dark_{D}} & =\frac{2\pi}{\hbar}\frac{1}{N_{D}}\sum_{l}\sum_{i,j}|\langle A_{l}|H_{DA}|D_{ij}\rangle|^{2}J_{A,D}-\frac{2\pi}{\hbar}\frac{1}{N_{D}}\sum_{l}\sum_{\vec{k}\in\text{FBZ}}|\langle A_{l}|H_{DA}|D_{\vec{k}}\rangle|^{2}J_{A,D}.
\end{align}
We note that the first term is exactly $\gamma_{\text{bare FRET}}'$.
Applying the arguments from the derivation (Section \ref{subsec:derivation_donor polaritons},
including the orientational averaging approximations of Eqs. (\ref{eq:approx separate})
and (\ref{eq:decouple})) of the rate of EET from donor polaritons
to acceptors, as well as the continuum approximation, we obtain
\begin{align}
\gamma_{A\leftarrow dark_{D}} & =\frac{2\pi}{\hbar}\rho_{A}^{(2D)}\mu_{D}^{2}\mu_{A}^{2}\langle\kappa_{FRET}^{2}\rangle\left[\frac{\int_{s}^{s+\mathcal{W}}dz\frac{\pi}{2(z_{0,A}-z)^{4}}}{\mathcal{W}}-\frac{1}{N_{D}}\sum_{\vec{k}\in\text{FBZ}}\frac{\int_{s}^{s+\mathcal{W}}dze^{-2a_{d\vec{k}}z}\frac{\pi}{2(z_{0,A}-z)^{4}}}{\int_{s}^{s+\mathcal{W}}dze^{-2a_{d\vec{k}}z}}\right]J_{AD}.
\end{align}
Next, we make the approximation $\frac{\int_{s}^{s+\mathcal{W}}dze^{-2a_{d\vec{k}}z}\frac{\pi}{2(z_{0,A}-z)^{4}}}{\int_{s}^{s+\mathcal{W}}dze^{-2a_{d\vec{k}}z}}\approx\frac{\int_{s}^{s+\mathcal{W}}dz\frac{\pi}{2(z_{0,A}-z)^{4}}}{\mathcal{W}}$,
where the lefthand side is a weighted average of $\frac{\pi}{2(z_{0,A}-z)^{4}}$.
Then we can write 
\begin{equation}
\gamma_{A\leftarrow dark_{D}}\approx\frac{2\pi}{\hbar}\rho_{A}^{(2D)}\mu_{D}^{2}\mu_{A}^{2}\langle\kappa_{FRET}^{2}\rangle\left(1-\frac{1}{N_{z,D}}\right)\frac{\int_{s}^{s+\mathcal{W}}dz\frac{\pi}{2(z_{0,A}-z)^{4}}}{\mathcal{W}}J_{AD},
\end{equation}
where we have used $\sum_{\vec{k}\in\text{FBZ}}=N_{xy,D}$ in obtaining
this equation. With $N_{z,D}\gg1$, we arrive at
\begin{align}
\gamma_{A\leftarrow dark_{D}} & \approx\frac{2\pi}{\hbar}\rho_{A}^{(2D)}\mu_{D}^{2}\mu_{A}^{2}\langle\kappa_{FRET}^{2}\rangle\frac{\frac{\pi}{6}\left[\frac{1}{(z_{0,A}-s-\mathcal{W})^{3}}-\frac{1}{(z_{0,A}-s)^{3}}\right]}{\mathcal{W}}J_{AD}.\label{eq:dark d to a_expl}
\end{align}
This expression is exactly the bare FRET rate $\gamma_{\text{bare FRET}}$
of Eq. (\ref{eq:last rate 1-1}) under the aforementioned assumptions
of infinitely extended slab along the $xy$ plane, translational symmetry,
orientational averaging, and the continuum limit. 

\subsection{Additional simulation notes/data for strongly coupling donor\label{subsec:additional simulations_sc donor}}

\begin{figure}
\includegraphics[width=1\textwidth]{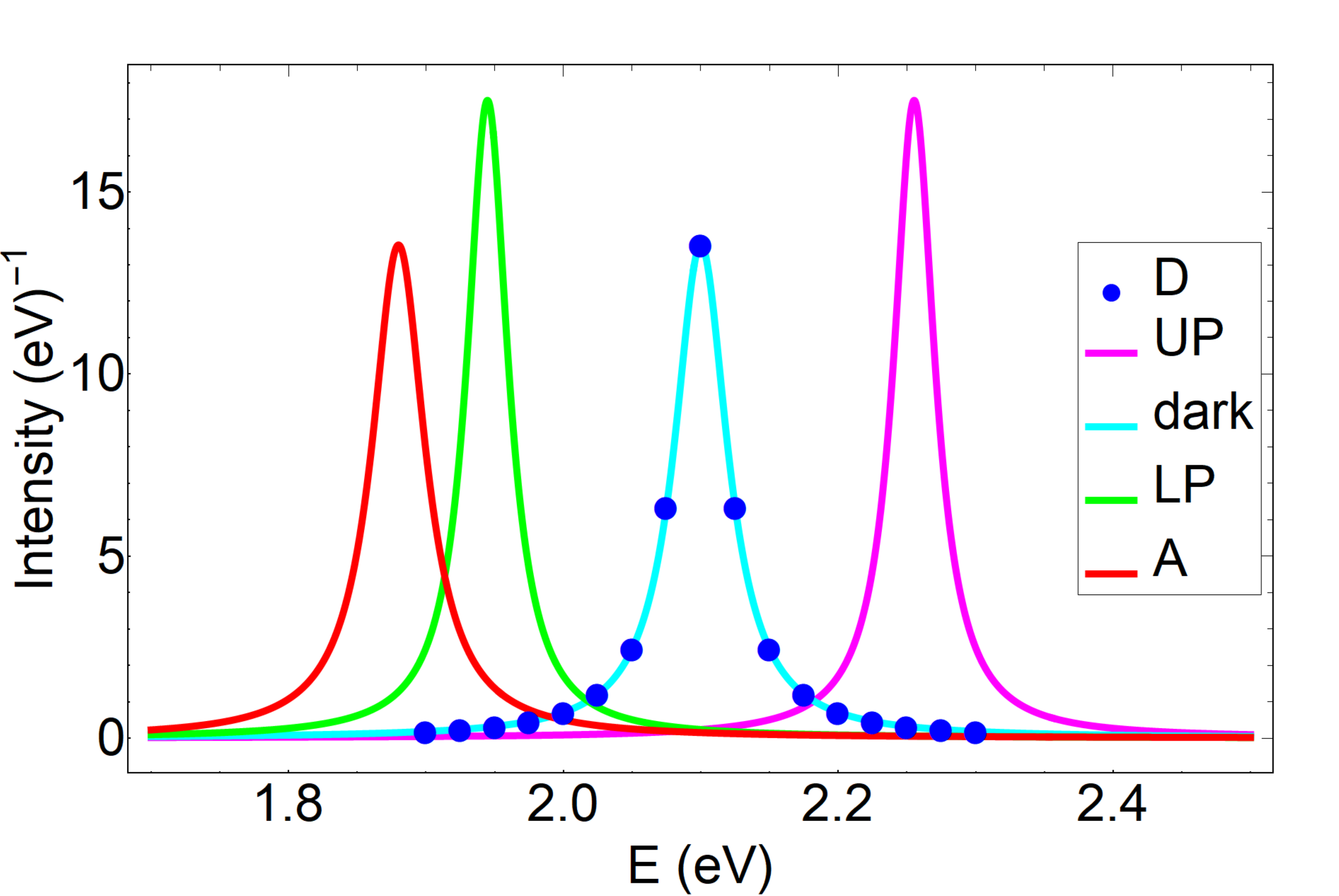}

\caption{SC of SPs to donors. Lorentzian spectra for the donor polariton and dark states, as well
as the bare donors and acceptors.}
\label{fgr:pd_overlaps} 
\end{figure}

\subsection{Derivation of simulated rates for strongly coupling acceptors\label{subsec:derivations sc acceptors}}

The simulated rates for EET from a monolayer of bare donor at $z=z_{0,D}>z_{m,A}$
for all $m$ to acceptor polariton/dark states asssuming a thick acceptor
slab ($N_{A}\gg N_{xy,A}$), orientational averaging and no correlations
for the TDMs $\vec{\mu}_{A_{lm}}$ are given by

\begin{subequations}\label{eq:rates_to_D_uncorr expl}
\begin{align}
\gamma_{\alpha_{A}\leftarrow D} & =\frac{1}{\hbar}\int_{0}^{\pi/a}dk\,k\left[|c_{A_{\vec{k}}\alpha_{A,\vec{k}}}|^{2}\mu_{D}^{2}\mu_{A}^{2}\langle\kappa_{FRET}^{2}\rangle\frac{\int_{s}^{s+\mathcal{W}}dze^{-2a_{d\vec{k}}z}\frac{\pi}{2(z_{0,D}-z)^{4}}}{\frac{e^{-2a_{d\vec{k}}(s+\mathcal{W})}-e^{-2a_{d\vec{k}}s}}{-2\alpha_{d\vec{k}}}}+|c_{\vec{k}\alpha_{A,\vec{k}}}|^{2}\mu_{D}^{2}\langle\kappa_{LM}^{2}\rangle\frac{\hbar\omega_{\vec{k}}}{2\epsilon_{0}L_{\vec{k}}}e^{-2a_{d\vec{k}}z_{0,D}}\right]J_{\alpha_{A,\vec{k}}D}\label{eq:d to pa_expl}\\
\gamma_{dark_{A}\leftarrow D} & =\frac{2\pi}{\hbar}\rho_{A}\mu_{D}^{2}\mu_{A}^{2}\langle\kappa_{FRET}^{2}\rangle\frac{\pi}{6}\left[\frac{1}{(z_{0,D}-s-\mathcal{W})^{3}}-\frac{1}{(z_{0,D}-s)^{3}}\right]J_{AD}=\gamma_{\text{bare FRET}}'.\label{eq:d to darka_expl}
\end{align}

\end{subequations}\noindent Here, the thickness of the acceptor slab
is $\mathcal{W}$, its base is located at $z=s$, and its (three-dimensional)
concentration is $\rho_{A}=\frac{N_{A}}{S\mathcal{W}}$. The derivation
of Eq. (\ref{eq:rates_to_D_uncorr expl}) starts with the preliminary
rate expression in Eq. (\ref{eq:rates_to_D}) of the main text and
proceeds analogously to those in Sections \ref{subsec:derivation_donor polaritons}
and \ref{subsec:derivation_dark donors}, respectively. In contrast
to Eq. (\ref{eq:plextoa}), Eq. (\ref{eq:d to pa_expl}) also sums
over the final polariton $\vec{k}$ modes, yielding a integral rate
upon invoking the continuum-limit transformation $\sum_{\vec{k}\in\text{FBZ}}\rightarrow\frac{S}{(2\pi)^{2}}\int_{0}^{2\pi}d\phi\int_{0}^{\pi/a}dk\,k$
for acceptor lattice spacing $a$. Eq. (\ref{eq:rates_to_D_uncorr expl})
is a more explicit form of Eqs. \ref{eq:rates_to_D_uncorr} in the
main text. 

We now consider the case when

\begin{equation}
\frac{1}{N_{xy,A}}\sum_{\vec{k}\in\text{FBZ}}|\langle\vec{k}|H_{DP}|D_{i0}\rangle|^{2}<\frac{1}{N_{A}}\sum_{l,m}|\langle A_{lm}|H_{DA}|D_{i0}\rangle|^{2},\label{eq:ineq}
\end{equation}
in other words, the average PRET coupling intensity is smaller than
that of FRET; this happens when the donor-acceptor separation lies
within the typical FRET range (1-10 nm). As we next show, the EET
rate from a monolayer of bare donors to acceptor polariton band $\alpha_{A}$
is consequently much smaller than the bare FRET rate ($\gamma_{\text{bare FRET}}''$
in Eq. (\ref{eq:bare FRET 3})) as $N_{A}\gg N_{xy,A}$ (\textit{i.e.},
$N_{z,A}\gg1$), for isotropic and uncorrelated orientational distribution
of TDMs $\vec{\mu}_{A_{ij}}$. The steps taken here resemble those
employed in Section \ref{subsec:derivation_dark donors}. Starting
from Eq. (\ref{eq:second rate k-1}), we utilize $|c_{A_{lm}A_{\vec{k}}}|^{2}\approx\frac{1}{N_{A}}$
to obtain
\begin{align}
\gamma_{\alpha_{A}\leftarrow D} & \approx\frac{2\pi}{\hbar N_{D}}\sum_{\vec{k}\in\text{FBZ}}\sum_{i}\left[|c_{A_{\vec{k}}\alpha_{A,\vec{k}}}|^{2}\sum_{l,m}\frac{1}{N_{A}}|\langle A_{lm}|H_{DA}|D_{i0}\rangle|^{2}+|c_{\vec{k}\alpha_{A,\vec{k}}}|^{2}|\langle\vec{k}|H_{DP}|D_{i0}\rangle|^{2}\right]J_{\alpha_{A,\vec{k}},D}\nonumber \\
 & \leq\frac{2\pi}{\hbar N_{D}}\sum_{i}\sum_{l,m}|\langle A_{lm}|H_{DA}|D_{i0}\rangle|^{2}\frac{1}{N_{z,A}}\max_{\vec{k}\in\text{FBZ}}J_{\alpha_{A,\vec{k}},D}\nonumber \\
 & \qquad+\frac{2\pi}{\hbar N_{D}}\sum_{i}\sum_{\vec{k}\in\text{FBZ}}|\langle\vec{k}|H_{DP}|D_{i0}\rangle|^{2}\max_{\vec{k}\in\text{FBZ}}J_{\alpha_{A,\vec{k}},D}\nonumber \\
 & \ll\frac{2\pi}{\hbar N_{D}}\sum_{i}\sum_{l,m}|\langle A_{lm}|H_{DA}|D_{i0}\rangle|^{2}J_{A,D}\nonumber \\
 & =\gamma_{\text{bare FRET}}'',
\end{align}
where the second inequality holds for sufficiently large $N_{z,A}$
such that $\frac{2}{N_{z,A}}\max_{\vec{k}\in\text{FBZ}}J_{\alpha_{A,\vec{k}},D}\ll J_{A,D}$.
This result can be physically interpreted as follows: $\gamma_{\text{bare FRET}}''$
and $\gamma_{\alpha_{A}\leftarrow D}^{FRET}$ scale as the number
of final states $N_{A}$ and $N_{xy,A}\ll N_{A}$, respectively.

\subsection{Additional simulation notes/data for strongly coupling acceptors\label{subsec:additional data_sc acceptors}}

We note that the rates from donors to polariton bands (Eq. \ref{eq:d to pa_expl})
were calculated numerically. In particular, the integrals over $k$
were calculated via the trapezoid rule using 2000 intervals.

\begin{figure}
\includegraphics[width=1\textwidth]{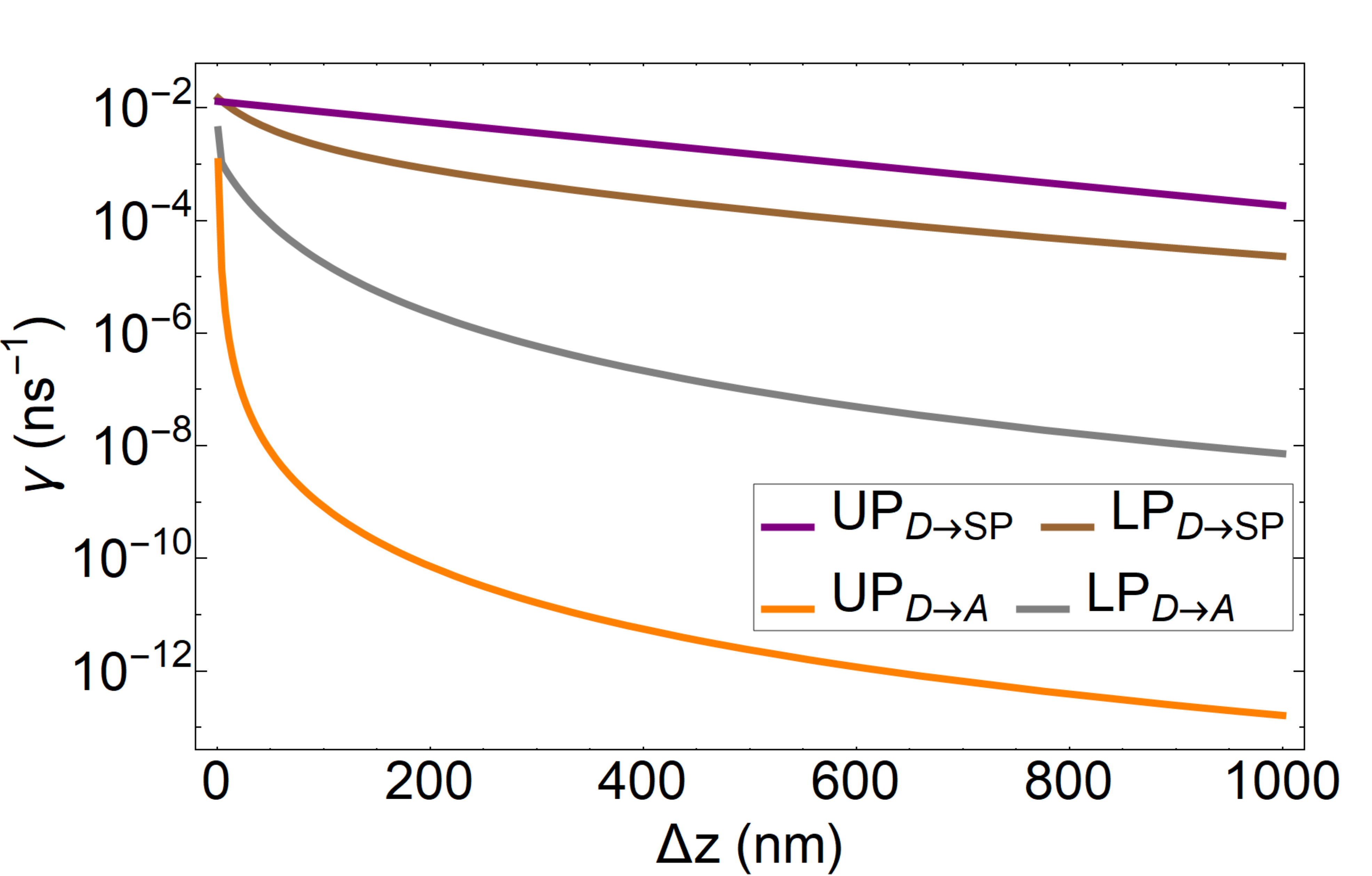}

\caption{SC of SPs to acceptors. Contributions of EET rates from donors to acceptor UP and LP due to
donor-acceptor and SP-donor interactions.}
\label{fgr:pa_breakdown} 
\end{figure}

\begin{figure}
\includegraphics{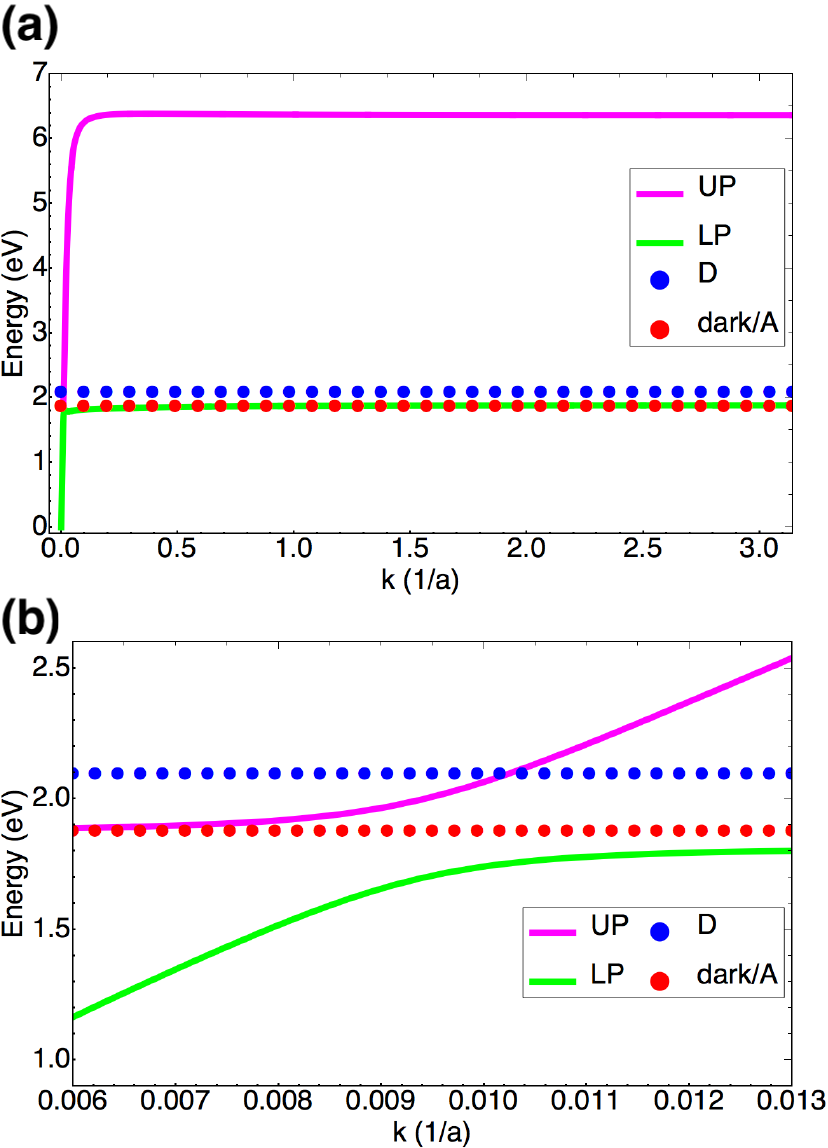}

\caption{SC of SPs to acceptors. (a) Dispersion curve in the FBZ. The value $a=1\times10^{-9}\text{ m}$ is the inter-chromophoric
lattice spacing for the acceptor slab. (b) Expanded view of dispersion
curve at region of anticrossing characteristic of SC.}
\label{fgr:pa_disp} 
\end{figure}

\subsection{Simulations for the ``carnival effect''\label{subsec:rates reversed}}

For this simulation, the coupling is strong enough such that the UP
is higher in energy than the bare donors. Thus, the acceptor UP becomes
a donor and the donors turn into acceptors. Thus, we use the rate
Eq. (\ref{eq:plextoa}) derived for the case of strongly-coupling
SP to donors only for transfer from a polariton state with the labels
$D$ and $A$ interchanged.

\begin{figure}
\includegraphics[width=1\textwidth]{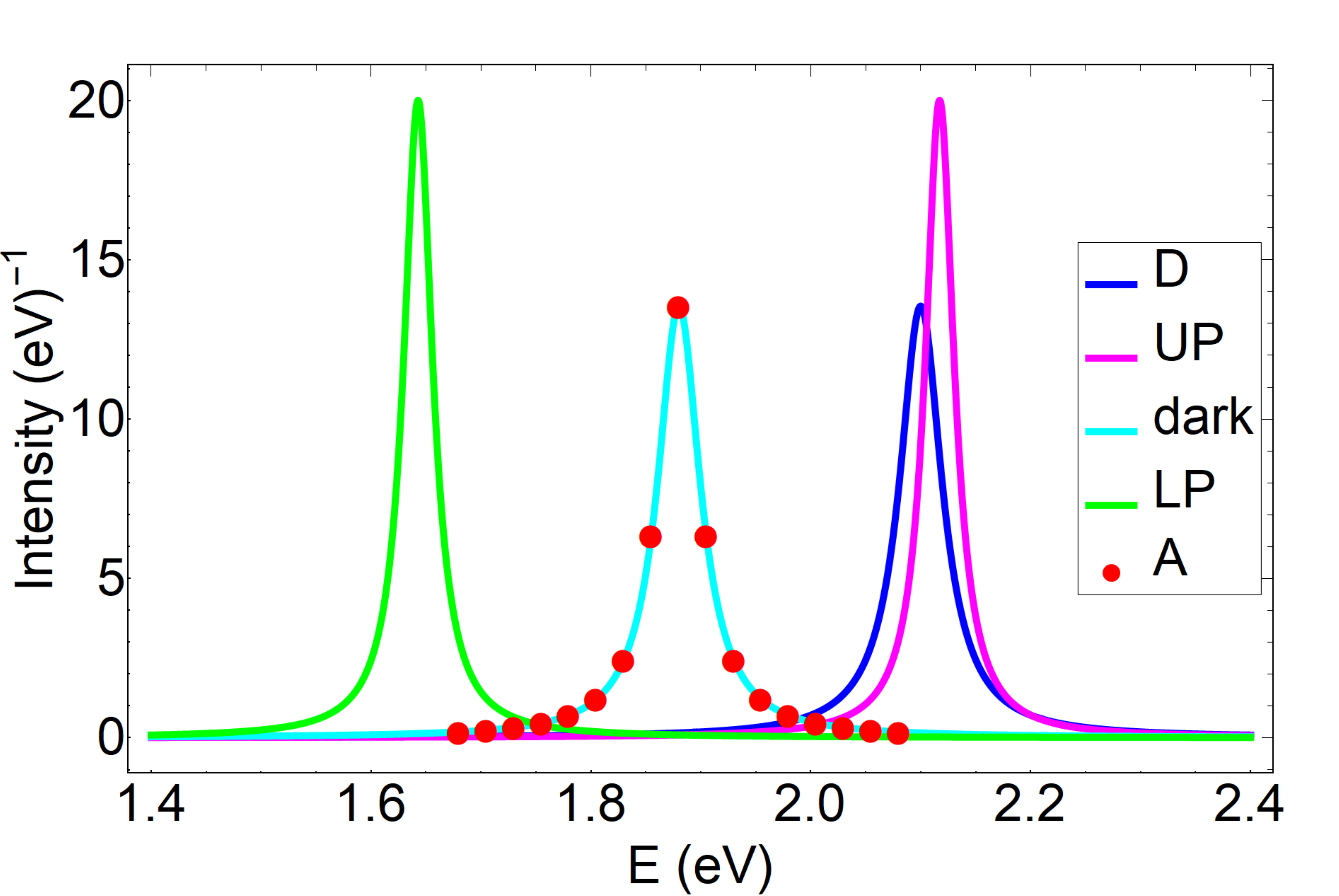}

\caption{Carnival effect or EET role-reversal. Lorentzian spectra for the acceptor
polariton and dark states, as well as the bare donors and acceptors.}
\label{fgr:pd_reverse_overlaps}
\end{figure}

\section{Case $ii$\label{sec:sc both} }

\subsection{Derivation: EET rates Eqs. (\ref{eq:rates_both}) \label{subsec:derivations_bb}}

In this section, we show how to derive the EET rates in Eqs. (\ref{eq:rates_both})
of the main text between polariton/dark states when both donors and
acceptors are strongly coupled to an SP mode. While we only derive
rate Eq. (\ref{eq:bright to bright}) in significant detail, Eqs.
(\ref{eq:from dark}) and (\ref{eq:last rate k}) can be analogously
obtained in essentially the same manner.

Excitons coupled to the $\vec{k}$-th SP mode produce polariton states
of the form $|\alpha_{\vec{k}}\rangle=c_{D_{\vec{k}}\alpha_{\vec{k}}}|D_{\vec{k}}\rangle+c_{A_{\vec{k}}\alpha_{\vec{k}}}|A_{\vec{k}}\rangle+c_{\vec{k}\alpha_{\vec{k}}}|\vec{k}\rangle$.
As discussed in the main text, the perturbation $V^{(ii)}$ due to
the chromophore and photon baths induces transitions between the UP,
MP, LP and dark states of donors and acceptors,
\begin{align}
V^{(ii)} & =\sum_{C}\sum_{i,j}|C_{ij}\rangle\langle C_{ij}|\sum_{q}\lambda_{q,C}\hbar\omega_{q,C}(b_{q,C_{ij}}^{\dagger}+\text{h.c.})+\sum_{\vec{k}}\sum_{q}g_{q,\vec{k}}(b_{q,P}^{\dagger}a_{\vec{k}}+\text{h.c.}),\label{eq:sys-B expl}
\end{align}
where $C=D,A$. Given the locality of vibronic coupling, the first
term only couples two states that share the same chromophores. On
the other hand, the second term does not couple different polariton/dark
states because it represents radiative and Ohmic losses from the SP
modes. The EET rate from a single polariton state $|\alpha_{\vec{k}}\rangle$
to the polariton band $\beta$ can then be calculated by invoking
Fermi's Golden Rule,
\begin{align}
\gamma_{\beta\leftarrow\alpha_{\vec{k}}} & =\sum_{\vec{k}'\in\text{FBZ}}\sum_{C}|c_{C_{\vec{k}'}\beta_{\vec{k}'}}|^{2}|c_{C_{\vec{k}}\alpha_{\vec{k}}}|^{2}\sum_{i,j}|c_{C_{ij}C_{\vec{k}'}}|^{2}|c_{C_{ij}C_{\vec{k}}}|^{2}\nonumber \\
 & \qquad\times\frac{2\pi}{\hbar^{2}}\sum_{L'}\sum_{L}p_{L_{C_{ij}}}|\langle L'_{C_{ij}}|\sum_{q}\lambda_{q,C}\hbar\omega_{q,C}(b_{q,C_{ij}}^{\dagger}+b_{q,C_{ij}})|L{}_{C_{ij}}\rangle|^{2}\delta(\omega_{\beta_{\vec{k}'}\alpha_{\vec{k}}}+\omega_{L'_{C_{ij}}L{}_{C_{ij}}}).
\end{align}
$|L_{C_{ij}}\rangle$ refers to a local bath state of molecule $C_{ij}$
and $p_{L_{C_{ij}}}$ is the thermal probability to populate it. Assuming
that all molecules of the same type $C$ (donor or acceptor) feature
the same bath modes, we can associate the single-molecule rate
\begin{align}
\frac{2\pi}{\hbar^{2}}\sum_{L'}\sum_{L}p_{L_{C_{ij}}}|\langle L'_{C_{ij}}|\sum_{q}\lambda_{q,C}\hbar\omega_{q,C}(b_{p,C_{ij}}^{\dagger}+b_{p,C_{ij}})|L_{C_{ij}}\rangle|^{2}\delta(\omega+\omega_{L'_{C_{ij}}L_{C_{ij}}}) & =\mathcal{R}_{C}(\omega)
\end{align}
which can be readily related to a spectral density, as explained in
the main text. Then we arrive at the compact expression,
\begin{equation}
\gamma_{\beta\leftarrow\alpha_{\vec{k}}}=\sum_{\vec{k}'\in\text{FBZ}}\sum_{C}|c_{C_{\vec{k}'}\beta_{\vec{k}'}}|^{2}|c_{C_{\vec{k}}\alpha_{\vec{k}}}|^{2}\sum_{i,j}|c_{C_{ij}C_{\vec{k}'}}|^{2}|c_{C_{ij}C_{\vec{k}}}|^{2}\mathcal{R}_{C}(\omega_{\beta_{\vec{k}'}\alpha_{\vec{k}}}),
\end{equation}
which is exactly Eq. (\ref{eq:bright to bright}) of the main text. 

To derive the average rate Eq. (\ref{eq:from dark}) from dark states
of chromophore $C$ to the polariton band $\alpha$, we begin with
\begin{equation}
\gamma_{\alpha\leftarrow dark_{C}}=\frac{2\pi}{\hbar^{2}}\sum_{\vec{k}'\in\text{FBZ}}\frac{1}{N_{C}-N_{xy,C}}\sum_{\vec{k}\in\text{FBZ}}\sum_{d}\sum_{i'j'}\sum_{L'}\sum_{i''j''}\sum_{L}p_{L_{C_{i''j''}}}|\langle\alpha_{\vec{k}'},L'_{C_{i'j'}}|V^{(ii)}|d_{C,\vec{k}},L{}_{C_{i''j''}}\rangle|^{2}\delta(\omega_{\alpha_{\vec{k}'}C}+\omega_{L'_{C_{i'j'}}L{}_{C_{i''j''}}}).\label{eq:pre from dark}
\end{equation}
Similarly, the derivation of rate Eq. (\ref{eq:last rate k}) from
polariton state $|\alpha_{\vec{k}}\rangle$ to the same dark states
starts at 
\begin{equation}
\gamma_{dark_{C}\leftarrow\alpha_{\vec{k}}}=\frac{2\pi}{\hbar^{2}}\sum_{\vec{k}\in\text{FBZ}}\sum_{d}\sum_{i'j'}\sum_{L'}\sum_{i''j''}\sum_{L}p_{L_{C_{i''j''}}}|\langle d_{C,\vec{k}'},L'_{C_{i'j'}}|V^{(ii)}|\alpha_{\vec{k}},L{}_{C_{i''j''}}\rangle|^{2}\delta(\omega_{C\alpha_{\vec{k}}}+\omega_{L'_{C_{i'j'}}L{}_{C_{i''j''}}}).\label{eq:pre last rate k}
\end{equation}
We note that the presence of the prefactor $\frac{1}{N_{C}-N_{xy,C}}$
in Eq. (\ref{eq:pre from dark}) and lack thereof in Eq. (\ref{eq:pre last rate k})
is due to the asymmetry of Fermi's Golden Rule: in the former equation,
dark states serve as the initial state and thus the term for each
state is weighted by its occupation probability, which we have assumed
to be uniform at $\frac{1}{N_{C}-N_{xy,C}}$ given their degeneracy.

\subsection{Simulated rates \label{subsec:rates_both}}

In this section, we present the rates used in our simulations for
both donors and acceptors strongly coupled to a SP assuming (as discussed
in the main text) orientationally averaged TDMs $\vec{\mu}_{C_{ij}}$,
$N_{xy,D}=N_{xy,A}=N_{xy}$, and $N_{z,C}\gg1$ for $C=D,A$. 

The overlap between $|C_{ij}\rangle$ and the collective mode $|C_{\vec{k}}\rangle$
can be written as $|c_{C_{ij}C_{\vec{k}}}|=\frac{\kappa_{\vec{k}C_{ij}}e^{-a_{d\vec{k}}z_{j,C}}}{\sqrt{\sum_{i,j}\kappa_{\vec{k}C_{ij}}^{2}e^{-2a_{d\vec{k}}z_{j,C}}}}$
and plugged into Eq. (\ref{eq:bright to bright}) to obtain the rate
for EET from polariton state $|\alpha_{\vec{k}}\rangle$ to band $\beta$:
\begin{equation}
\gamma_{\beta\leftarrow\alpha_{\vec{k}}}=\frac{1}{2\pi}\sum_{C}\frac{|c_{C_{\vec{k}}\alpha}|^{2}}{\rho_{C}\frac{e^{-2a_{d\vec{k}}(s_{C}+\mathcal{W}_{C})}-e^{-2a_{d\vec{k}}s_{C}}}{-2a_{d\vec{k}}}}\int_{0}^{\pi/a}dk'\,k'|c_{C_{\vec{k}}\beta}|^{2}\frac{\frac{e^{-2(a_{d\vec{k}}+a_{d\vec{k}'})(s_{D}+\mathcal{W}_{D})}-e^{-2(a_{d\vec{k}}+a_{d\vec{k}'})s_{D}}}{-4a_{d\vec{k}'}}}{\frac{e^{-2a_{d\vec{k}'}(s_{D}+\mathcal{W}_{D})}-e^{-2a_{d\vec{k}'}s_{D}}}{-2a_{d\vec{k}'}}}\mathcal{R}_{D}(\omega_{\beta_{\vec{k}'}\alpha_{\vec{k}}}),
\end{equation}
where we have applied orientational averaging of the TDMs under the
approximation $\langle|c_{C_{ij}C_{\vec{k}}}|^{2}\rangle\approx\frac{e^{-2a_{d\vec{k}}z_{j,C}}}{N_{xy}\sum_{j}e^{-2a_{d\vec{k}}z_{j,C}}}$
(Section \ref{subsec:derivation_donor polaritons}) and the $\vec{k}$-space
continuum-limit transformation (Section \ref{subsec:derivations sc acceptors}). 

The derivation of the EET rate from polariton $|\alpha_{\vec{k}}\rangle$
to dark states is quite alike that of Section \ref{subsec:derivation_dark donors}.
We can write Eq. (\ref{eq:last rate k}) as
\begin{align}
\gamma_{dark_{C}\leftarrow\alpha_{\vec{k}}} & =|c_{C_{\vec{k}}\alpha_{\vec{k}}}|^{2}\mathcal{R}_{C}(\omega_{C\alpha_{\vec{k}}})\sum_{i,j}|c_{C_{ij}C_{\vec{k}}}|^{2}\sum_{\vec{k}'\in\text{FBZ}}\sum_{d}|c_{C_{\vec{k}'}d_{C,\vec{k}'}}|^{2}|c_{C_{ij}C_{\vec{k}'}}|^{2}\nonumber \\
 & =|c_{C_{\vec{k}}\alpha_{\vec{k}}}|^{2}\mathcal{R}_{C}(\omega_{C\alpha_{\vec{k}}})\sum_{i,j}|c_{C_{ij}C_{\vec{k}}}|^{2}\sum_{\vec{k}'\in\text{FBZ}}\sum_{d}|\langle C_{ij}|d_{C,\vec{k}'}\rangle|^{2}.
\end{align}
Inserting the relation $\sum_{\vec{k}\in\text{FBZ}}\sum_{d}|d_{C,\vec{k}}\rangle\langle d_{C,\vec{k}}|=1_{C}^{(sys)}-\sum_{\vec{k}\in\text{FBZ}}|C_{\vec{k}}\rangle\langle C_{\vec{k}}|$,
we obtain
\begin{align}
\gamma_{dark_{C}\leftarrow\alpha_{\vec{k}}} & =|c_{C_{\vec{k}}\alpha_{\vec{k}}}|^{2}\mathcal{R}_{C}(\omega_{C\alpha_{\vec{k}}})\sum_{i,j}|c_{C_{ij}C_{\vec{k}}}|^{2}\left(1-\sum_{\vec{k}'\in\text{FBZ}}|c_{C_{ij}C_{\vec{k}'}}|^{2}\right).
\end{align}
Noting that $|c_{C_{ij}C_{\vec{k}'}}|^{2}\sim\frac{1}{N_{xy}N_{z,C}}$
and $\sum_{\vec{k}\in\text{FBZ}}$ sums over $N_{xy}$ terms, we utilize
$N_{z,C}\gg1$ and $\sum_{i,j}|c_{C_{ij}C_{\vec{k}}}|^{2}=1$ to write
\begin{equation}
\gamma_{dark_{C}\leftarrow\alpha_{\vec{k}}}=|c_{C_{\vec{k}}\alpha_{\vec{k}}}|^{2}\mathcal{R}_{C}(\omega_{C\alpha_{\vec{k}}}).
\end{equation}
 This same logic can be applied to arrive at 
\begin{equation}
\gamma_{\alpha\leftarrow dark_{C}}=\frac{1}{2\pi n_{C}\mathcal{W}_{C}}\int_{0}^{\pi/a}dk\,k|c_{C_{\vec{k}'}\alpha_{\vec{k}'}}|^{2}\mathcal{R}_{C}(\omega_{\alpha_{\vec{k}'}C})
\end{equation}
for a continuum of $\vec{k}$ states. 

In summary, the simulated equations for the case where both donors
and acceptors are strongly coupled are 

\begin{subequations}\label{eq:both expl}
\begin{align}
\gamma_{\beta\leftarrow\alpha_{\vec{k}}} & =\frac{1}{2\pi}\sum_{C}\frac{|c_{C_{\vec{k}}\alpha}|^{2}}{\rho_{C}\frac{e^{-2a_{d\vec{k}}(s_{C}+\mathcal{W}_{C})}-e^{-2a_{d\vec{k}}s_{C}}}{-2a_{d\vec{k}}}}\int_{0}^{\pi/a}dk'\,k'|c_{C_{\vec{k}}\beta}|^{2}\frac{\frac{e^{-2(a_{d\vec{k}}+a_{d\vec{k}'})(s_{D}+\mathcal{W}_{D})}-e^{-2(a_{d\vec{k}}+a_{d\vec{k}'})s_{D}}}{-4a_{d\vec{k}'}}}{\frac{e^{-2a_{d\vec{k}'}(s_{D}+\mathcal{W}_{D})}-e^{-2a_{d\vec{k}'}s_{D}}}{-2a_{d\vec{k}'}}}\mathcal{R}_{D}(\omega_{\beta_{\vec{k}'}\alpha_{\vec{k}}}),\label{eq:bb expl}\\
\gamma_{\alpha\leftarrow dark_{C}} & =\frac{1}{2\pi\rho_{C}\mathcal{W}_{C}}\int_{0}^{\pi/a}dk\,k|c_{C_{\vec{k}'}\alpha_{\vec{k}'}}|^{2}\mathcal{R}_{C}(\omega_{\alpha_{\vec{k}'}C}),\label{eq:db}\\
\gamma_{dark_{C}\leftarrow\alpha_{\vec{k}}} & =|c_{C_{\vec{k}}\alpha_{\vec{k}}}|^{2}\mathcal{R}_{C}(\omega_{C\alpha_{\vec{k}}}).\label{eq:bd}
\end{align}

\end{subequations}\noindent From these expressions, it can be seen
that the rates to polariton bands (Eqs. (\ref{eq:bb expl}) and (\ref{eq:db}))
scale as $\sim\frac{1}{N_{z,C}}$ relative to the rate to dark states
(Eq. (\ref{eq:bd})), which scales as a single-molecule rate $\mathcal{R}_{C}$.
This is because there are $N_{xy}$ states in each polariton band
as opposed to $N_{C}-N_{xy}$ in the band of dark states.

As discussed in the main text (Section \ref{sec:application}), the
spectral density that governs $\mathcal{R}_{C}(\omega)$ depends on
the energies $\hbar\omega_{q}$ and coupling strengths $\lambda_{q}\hbar\omega_{q}$
of representative localized vibrational modes $B_{\text{TDBC}}$,
reproduced below from literature:\cite{Coles2011}

\begin{tabular}{|c|>{\centering}p{1cm}|>{\centering}p{1cm}|>{\centering}p{1cm}|>{\centering}p{1cm}|>{\centering}p{1cm}|>{\centering}p{1cm}|}
\hline 
$\hbar\omega_{q}$ (meV) & 40 & 80 & 120 & 150 & 185 & 197\tabularnewline
\hline 
$\lambda_{q}\hbar\omega_{q}$ (meV) & 14 & 18 & 25 & 43 & 42 & 67\tabularnewline
\hline 
\end{tabular}

\subsection{Additional simulation notes/data\label{subsec:additional_res d_fix d}}

We note that the rates in Eqs. (\ref{eq:bb expl}) and (\ref{eq:db})
from polariton and dark states, respectively, to polariton bands were
calculated numerically: as for the case of strongly coupling only
the acceptors to SPs (Section \ref{subsec:additional data_sc acceptors}),
the $k$-integrals were computed with the trapezoid rule using 2000
intervals.

\begin{figure}
\includegraphics[width=1\textwidth]{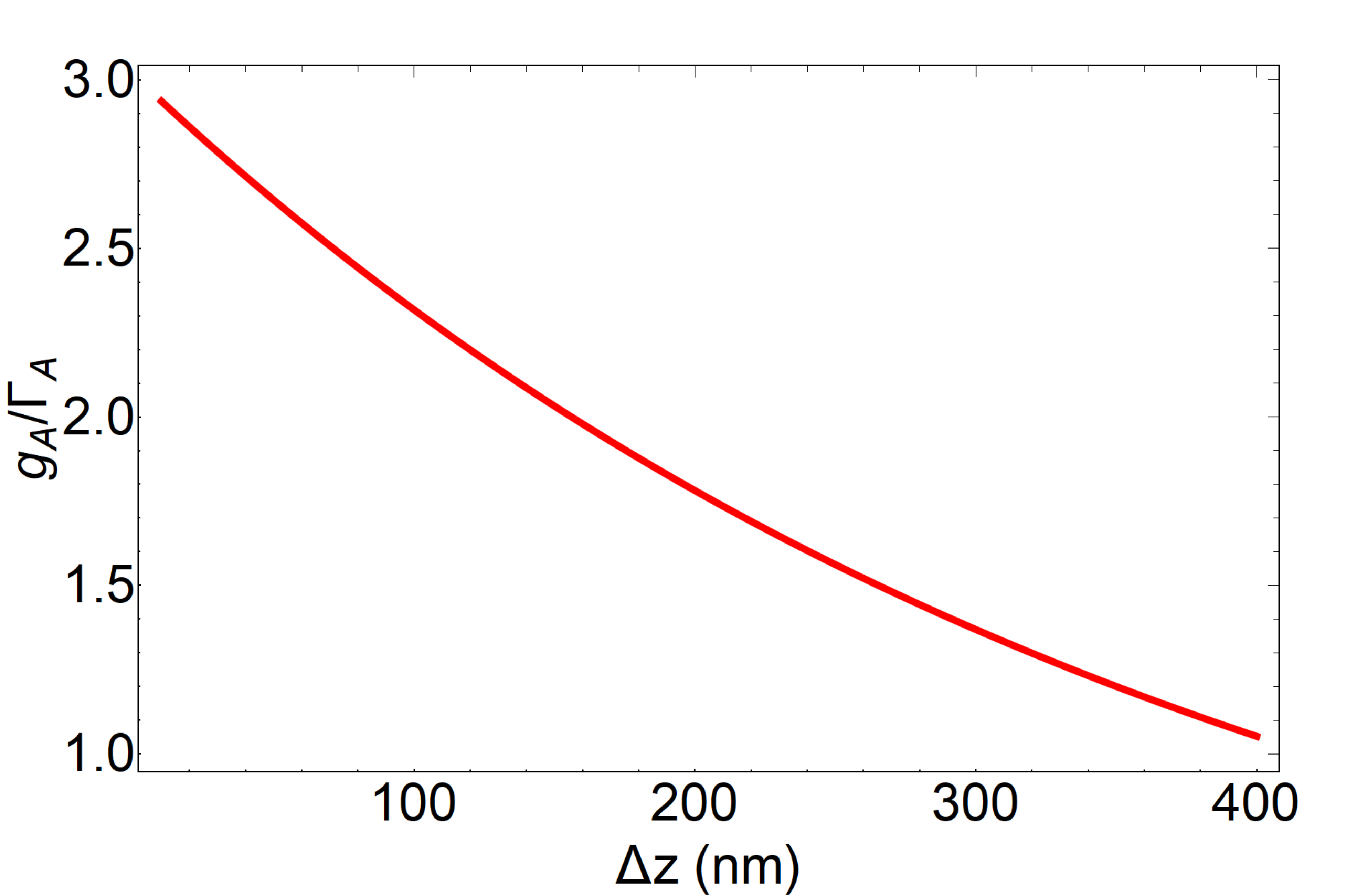}

\caption{SC of SPs to both donors and acceptors. Ratio of SP collective coupling
and linewidth of acceptors as a function of donor-acceptor separation
$\Delta z$. }
\label{fgr:pda_d_fixd_coupling}
\end{figure}

\begin{figure}
\includegraphics[width=1\textwidth]{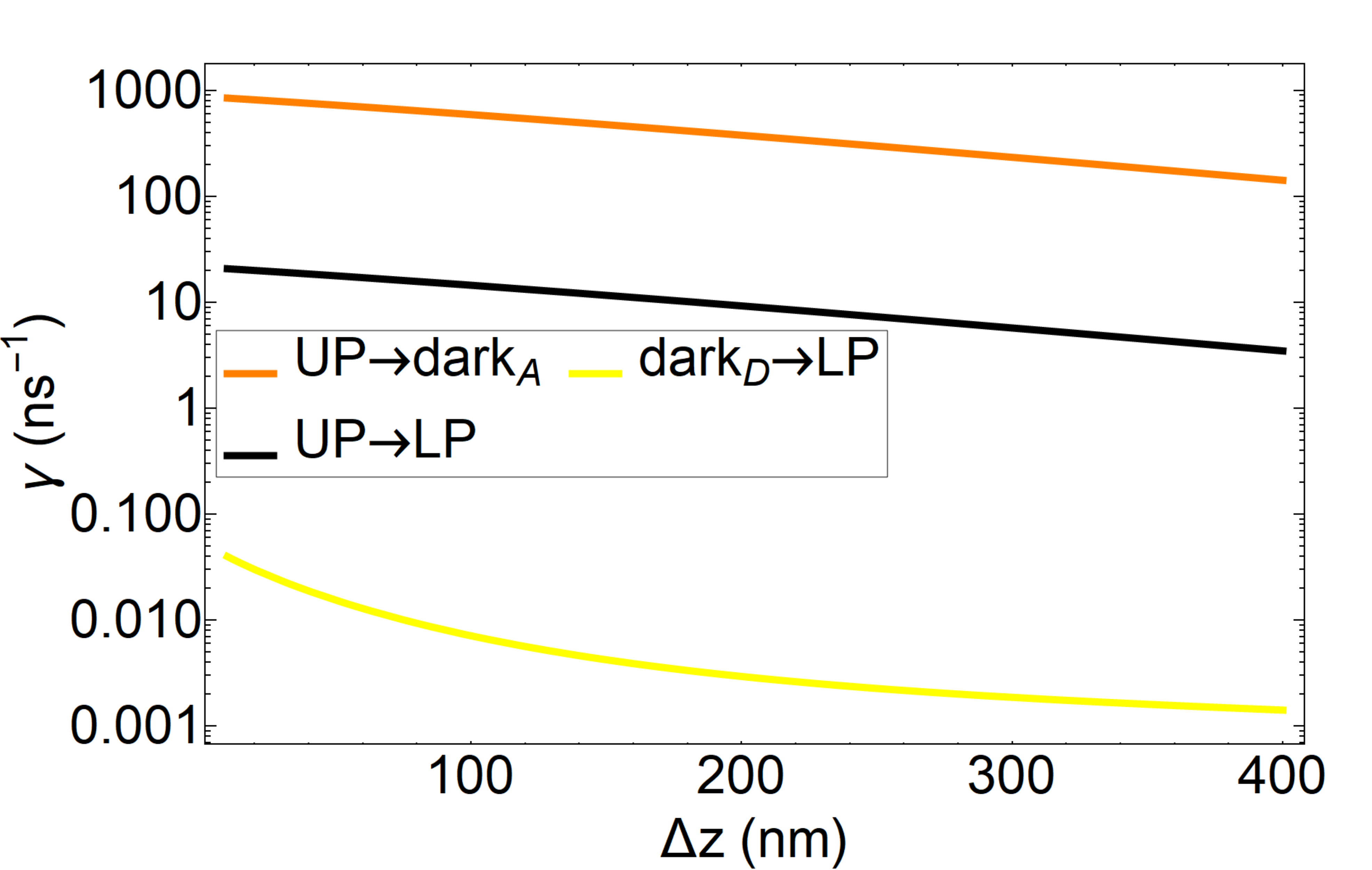}

\caption{SC of SPs to both donors and acceptors. EET rates (that are not shown
in Fig. \ref{fgr:pda_d_fixd} of the main text) as a function of donor-acceptor
separation $\Delta z$ for downhill transtions among polariton and
dark states. for The SP mode is resonant with the donor, and the donor
slab lies 1 nm from the metal and has fixed position while the acceptor
slab is moved in the $+z$-direction to vary $\Delta z$.}
\label{fgr:pda_d_fixd_minor} 
\end{figure}

\begin{figure}
\includegraphics{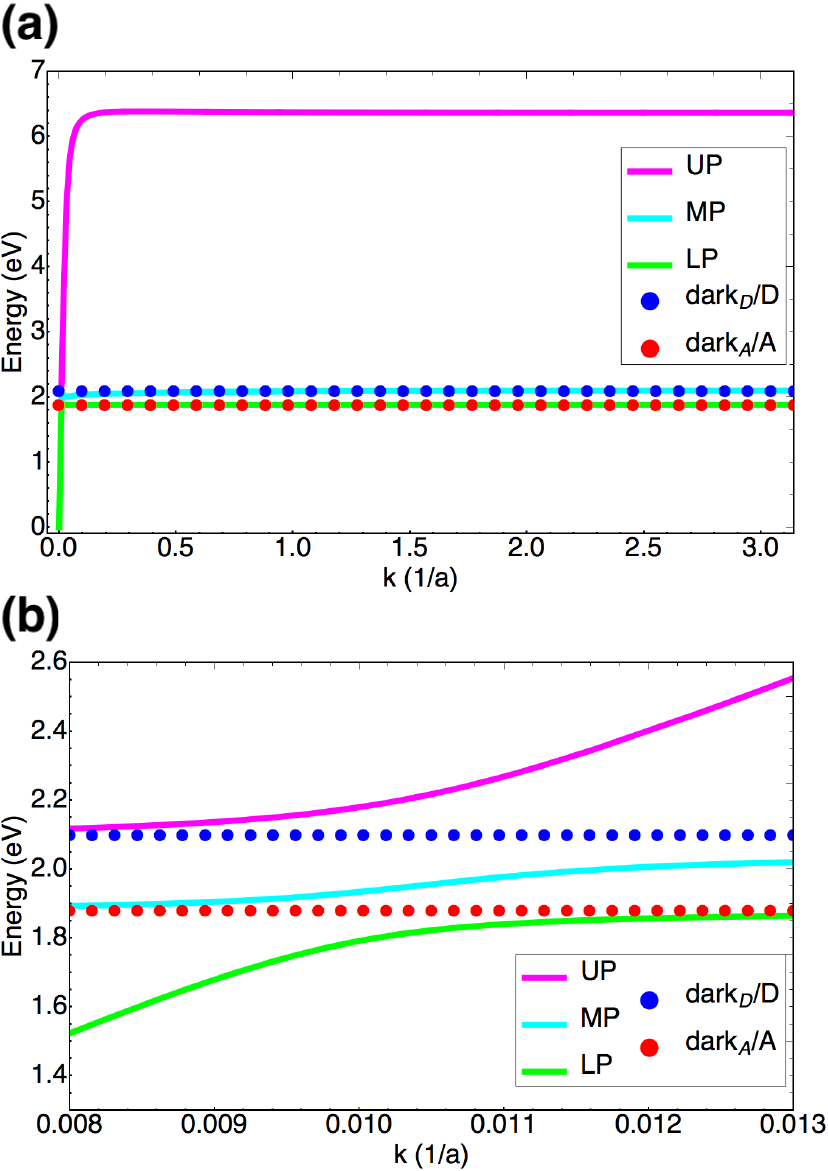}

\caption{SC of SPs to both donors and acceptors. (a) Dispersion curve in the FBZ. Donor-acceptor separation is 205 nm. The dispersion curves
at all other donor-acceptor separations (10-400 nm) have the same
qualitative form. The value $a=1\times10^{-9}\text{ m}$ is the inter-chromophoric
lattice spacing for both donor and acceptor slabs. (b) Expanded view
of dispersion curve at region of anticrossing characteristic of SC.}
\label{fgr:pda_d_fixd_disp} 
\end{figure}

\begin{figure}
\includegraphics{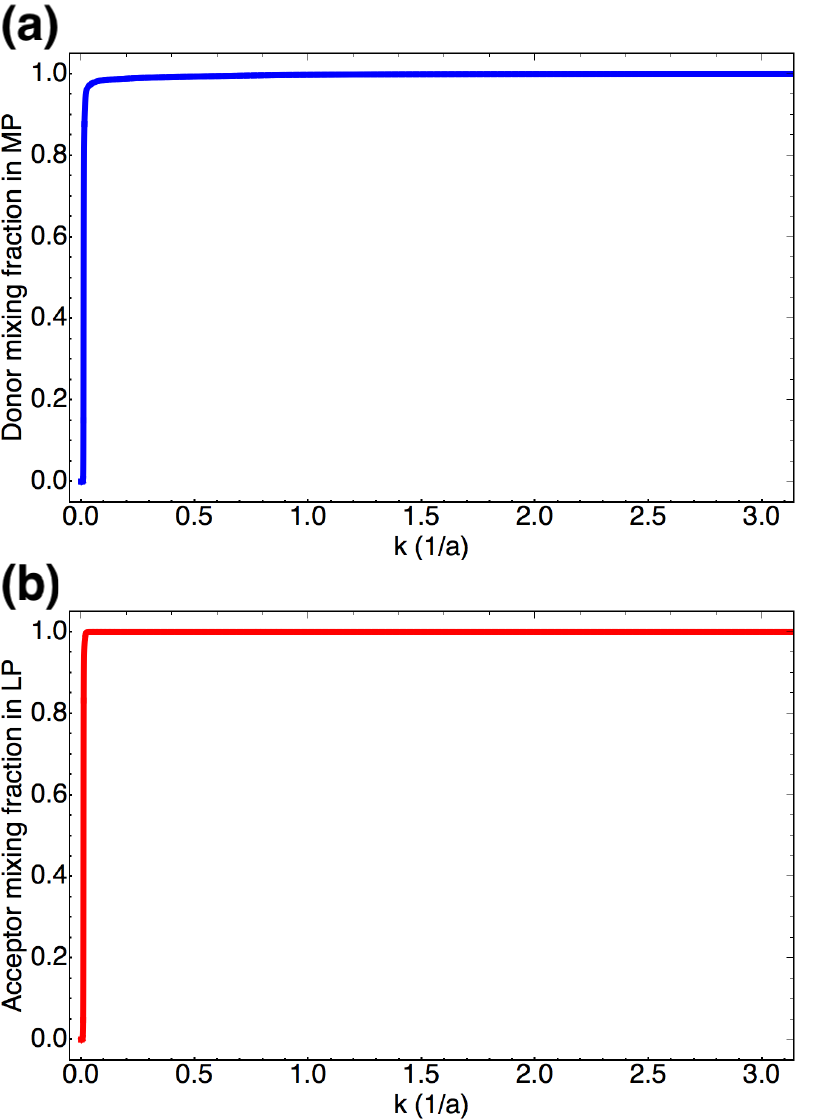}

\caption{SC of SPs to both donors and acceptors. (a) Donor mixing fraction $|c_{D_{\vec{k}}\text{MP}_{\vec{k}}}|^{2}$
in MP and (b) acceptor mixing fraction $|c_{A_{\vec{k}}\text{LP}_{\vec{k}}}|^{2}$
in LP. Donor-acceptor
separation is 205 nm. The mixing-fraction curves as a function of
$\vec{k}\in\text{FBZ}$ at all other donor-acceptor separations have
the same qualitative form. The value $a=1\times10^{-9}\text{ m}$
is the inter-chromophoric lattice spacing for both donor and acceptor
slabs. }
\label{fgr:pda_d_fixd_char} 
\end{figure}

\end{document}